\documentclass[10pt]{article}
\pdfoutput=1
\usepackage{jheppub}
\usepackage{amsmath,amssymb,amsthm,latexsym,bbm,calc,wasysym,mathtools, bbold}
\usepackage[english]{babel}
\usepackage{graphicx,color}
\usepackage[dvipsnames]{xcolor}
\usepackage{tikz-cd}
\usepackage{tikz}
\usetikzlibrary{shapes.geometric}
\usetikzlibrary{arrows,matrix,calc,scopes,decorations.markings}
\usepackage{xspace}
\usepackage{pifont,dsfont}
\usepackage{marvosym}
\usepackage{slashed}
\usepackage{subfigure}
\usepackage{sidecap}

\usepackage[export]{adjustbox}
\mathtoolsset{showonlyrefs=false}
\usepackage{enumitem}

\def\be{\begin{equation}}
\def\ee{\end{equation}}
\newcommand\sss{\scriptstyle}
\newcommand{\q}{\quad}
\newcommand{\mG}{\mathcal{G}}
\newcommand{\mD}{\mathcal{D}}

\newcommand{\wro}{\widetilde{\rho}}
\newcommand{\la}{\langle}
\newcommand{\ra}{\rangle}
\newcommand{\nn	}{\nonumber}
\newcommand{\tor}{\mathbb{T}_2}
\newcommand{\hee}{\Sigma_\mathsf{H}(\triangle)}

\newcommand{\blow}{\pmb{\large\triangle}_1}
\newcommand{\smo}{\,{\sss \otimes}\,}
\newcommand{\unitG}{\mathbbm{1}_G}
\newcommand{\unit}{\mathbbm{1}}
\newcommand{\cyl}{\mathbb{I}}

\def\ba{\begin{eqnarray}}
\def\ea{\end{eqnarray}}

\def\SU{{\rm{SU}}}

\definecolor{specBlue}{RGB}{113,119,165}

\def\k{{\rm k}}

\hypersetup{
	colorlinks=true,       
	linkcolor=BlueViolet,          
	citecolor=BrickRed,        
	filecolor=BrickRed,      
	urlcolor=BrickRed           
}

\tikzset{->-/.style={decoration={
  markings,
  mark=at position .5 with {\arrow{>}}},postaction={decorate}}}
\tikzset{-<-/.style={decoration={
  markings,
  mark=at position .5 with {\arrow{<}}},postaction={decorate}}}

\newcommand{\fourTet}[8]{
	\begin{tikzpicture}[scale=#7,baseline]
	\coordinate (a) at(0,0);    
	\coordinate (b) at (0.8,-0.75);  
	\coordinate (c) at (2.5,0);      
	\coordinate (d) at (1.25,2);     
	\coordinate (e) at ($ (a) ! 0.3 ! (c) $);
	\coordinate (f) at ($ (a) ! 0.45 ! (c) $);
	\coordinate (g) at ($ (a) ! 0.5 ! (b) $);
	\coordinate (h) at ($ (c) ! 0.5 ! (d) $);
	\coordinate (n) at ($ (g) ! 0.5 ! (h) $); 
	\draw (a) -- (b) -- (c) -- (d) -- cycle;
	\draw (b) -- (d);
	\draw (a) -- (e);
	\draw (f) -- (c);
	\ifnum #8=1 {
		\draw[-<,>=latex,line width=0.01pt] (a) -- ($ (a) ! 0.5 ! (b) $);
		\draw[-<,>=latex,line width=0.01pt] (b) -- ($ (b) ! 0.5 ! (c) $);
		\draw[-<,>=latex,line width=0.01pt] (c) -- ($ (c) ! 0.5 ! (d) $);
		\draw[<-,>=latex,line width=0.01pt] (f) -- (c);
		\draw[-<,>=latex,line width=0.01pt] (b) -- ($ (b) ! 0.5 ! (d) $);
		\draw[-<,>=latex,line width=0.01pt] (a) -- ($ (a) ! 0.5 ! (d) $);}
	\fi
	
	\node[left] at(a) {\scalebox{0.7}{$#1$}};
	\node[below] at(b) {\scalebox{0.7}{$#2$}};
	\node[right] at(c) {\scalebox{0.7}{$#4$}};
	\node[above] at(d) {\scalebox{0.7}{$#3$}};
	\ifnum #5>0 {
		\node[circle,fill=black,outer sep=0pt, inner sep=0.6pt] at(n) {};
		\draw[dashed] (a) -- (n);
		\draw[dashed] (b) -- (n);
		\draw[dashed] (c) -- (n);
		\draw[dashed] (d) -- (n);
	}\fi
	
	\ifnum #5=1 {
		\node[right, outer sep=0pt, inner sep=1pt] at($ (n) ! 0.07 ! (d) $) {\scalebox{0.7}{$#6$}};
		\ifnum #8=1 {
			\draw[->,>=latex,line width=0.01pt] ($ (a) ! 0.4 ! (n) $) -- ($ (a) ! 0.5 ! (n) $) ;
			\draw[->,>=latex,line width=0.01pt] ($ (b) ! 0.4 ! (n) $) -- ($ (b) ! 0.5 ! (n) $) ;
			\draw[->,>=latex,line width=0.01pt] ($ (c) ! 0.4 ! (n) $) -- ($ (c) ! 0.5 ! (n) $) ;
			\draw[->,>=latex,line width=0.01pt] ($ (d) ! 0.4 ! (n) $) -- ($ (d) ! 0.5 ! (n) $) ;}\fi}
	\fi
	\ifnum #5=2 {
		\node[right, outer sep=0pt, inner sep=1pt] at($ (n) ! 0.07 ! (d) $) {\scalebox{0.7}{$#6$}};
		\ifnum #8=1 {
			\draw[-<,>=latex,line width=0.01pt] ($ (a) ! 0.4 ! (n) $) -- ($ (a) ! 0.5 ! (n) $) ;
			\draw[->,>=latex,line width=0.01pt] ($ (b) ! 0.4 ! (n) $) -- ($ (b) ! 0.5 ! (n) $) ;
			\draw[->,>=latex,line width=0.01pt] ($ (c) ! 0.4 ! (n) $) -- ($ (c) ! 0.5 ! (n) $) ;
			\draw[->,>=latex,line width=0.01pt] ($ (d) ! 0.4 ! (n) $) -- ($ (d) ! 0.5 ! (n) $) ;}\fi}
	\fi
	\ifnum #5=3 {
		\node[right, outer sep=0pt, inner sep=1pt] at($ (n) ! 0.07 ! (d) $) {\scalebox{0.7}{$#6$}};
		\ifnum #8=1 {
			\draw[-<,>=latex,line width=0.01pt] ($ (a) ! 0.4 ! (n) $) -- ($ (a) ! 0.5 ! (n) $) ;
			\draw[-<,>=latex,line width=0.01pt] ($ (b) ! 0.4 ! (n) $) -- ($ (b) ! 0.5 ! (n) $) ;
			\draw[->,>=latex,line width=0.01pt] ($ (c) ! 0.4 ! (n) $) -- ($ (c) ! 0.5 ! (n) $) ;
			\draw[->,>=latex,line width=0.01pt] ($ (d) ! 0.4 ! (n) $) -- ($ (d) ! 0.5 ! (n) $) ;}\fi}
	\fi
	\ifnum #5=4 {
		\node[right, outer sep=0pt, inner sep=1pt] at($ (n) ! 0.07 ! (d) $) {\scalebox{0.7}{$#6$}};
		\ifnum #8=1 {
			\draw[-<,>=latex,line width=0.01pt] ($ (a) ! 0.4 ! (n) $) -- ($ (a) ! 0.5 ! (n) $) ;
			\draw[-<,>=latex,line width=0.01pt] ($ (b) ! 0.4 ! (n) $) -- ($ (b) ! 0.5 ! (n) $) ;
			\draw[-<,>=latex,line width=0.01pt] ($ (c) ! 0.4 ! (n) $) -- ($ (c) ! 0.5 ! (n) $) ;
			\draw[->,>=latex,line width=0.01pt] ($ (d) ! 0.4 ! (n) $) -- ($ (d) ! 0.5 ! (n) $) ;}\fi}
	\fi
	\end{tikzpicture}
}

\title{\boldmath Towards a dual spin network basis for (3+1)d lattice gauge theories and topological phases}

\author[a,b]{Clement Delcamp,}
\author[a]{Bianca Dittrich}

\affiliation[a]{Perimeter Institute for Theoretical Physics,\\ 31 Caroline Street North, Waterloo, Ontario  N2L 2Y5, Canada}
\affiliation[b]{Department of Physics $\&$ Astronomy and Guelph-Waterloo Physics Institute \\  University of Waterloo, Waterloo, Ontario N2L 3G1, Canada}

\emailAdd{cdelcamp@perimeterinstitute.ca}
\emailAdd{bdittrich@perimeterinstitute.ca}

\abstract{Using a recent strategy to encode the space of flat connections on a three-manifold with string-like defects into the space of flat connections on a so-called 2d Heegaard surface, we propose a novel way to define gauge invariant bases for (3+1)d lattice gauge theories and gauge models of topological phases. In particular, this method reconstructs the spin network basis and yields a novel dual spin network basis. While the spin network basis allows to interpret states in terms of electric excitations, on top of a vacuum sharply peaked on a vanishing electric field, the dual spin network basis describes magnetic (or curvature) excitations,  on top of a vacuum sharply peaked on a vanishing magnetic field (or flat connection). This technique is also applicable for manifolds with boundaries. We distinguish in particular a dual pair of boundary conditions, namely of electric type and of magnetic type. This can be used to consider a generalization of Ocneanu's tube algebra in order to reveal the algebraic structure of the excitations associated with certain 3d manifolds.}

\begin{document} 
	\vspace*{-2em}
	\maketitle
	\flushbottom

\newpage

\section{Introduction}

\emph{Gauge theories} have become indispensable in modern physics. A prime example of gauge theories is Yang-Mills theory, which is an essential part of the standard model for particle physics. A crucial tool to address non-perturbative regimes is \emph{lattice gauge theory}, which relies upon a discretization of spacetime. Lattice gauge theories appear, in particular, in the description of quantum chromodynamics and \emph{loop quantum gravity} \cite{rovelli2004, thiemann, Perez:2004hj}. They also provide a particularly tractable class of \emph{topological quantum field theories} which in turn describe gapped phases of matter \cite{fradkin2013field, wen2004quantum}. An example of such topological quantum field theory is $BF$ theory that can be understood as the weak coupling limit of Yang-Mills theory. It is an important ingredient of loop quantum gravity \cite{Perez:2004hj, DGflux, DGfluxC, DGfluxQ,Lewandowski:2015xqa} as well as the spin foam approach \cite{Baez:1997zt, Perez:2012wv}. In the condensed matter literature, its Hamiltonian realization for finite groups is known as the Kitaev model. 

Via the introduction of gauge variables, gauge theories allow for a local formulation of their dynamics---at the price of introducing redundancies. In contrast, gauge invariant observables feature a form of non-locality. In lattice gauge theories, observables are typically defined over extended objects. Indeed, holonomies arise from a (exponentiated) connection integrated over a path. Choosing this path to be a loop and taking the trace of the holonomy, we obtain a gauge invariant observable known as a \emph{Wilson loop}. Conjugated to these holonomies are electrical fluxes which are integrals of the electrical field over surfaces (in three spatial dimensions).\footnote{In non-abelian theories this electrical field is parallel transported to a common frame before integration.} A key challenge of such formulation is to define a  set of commuting gauge invariant observables such that the eigenvalues of these observables provide a complete---but not over-complete---labeling for a basis of gauge invariant states. Additionally, one would like these observables to be as local as possible. This is needed in order to define a notion of (quasi-)local excitations which arise as defects in the context of topological phases.

The \emph{spin network basis} \cite{Rovelli:1995ac} provides a gauge invariant basis which diagonalizes certain invariant combinations of electrical fluxes. This basis has found wide applications in loop quantum gravity \cite{Rovelli:1994ge,rovelli2004, thiemann, Perez:2004hj} and more generally in lattice gauge theories. It also appears in the form of \emph{string nets} \cite{Levin2004} in the study of (2+1)d topological phases of matter. In spite of its success, the spin network basis is mostly adjusted to describe the strong coupling regime of lattice gauge theories. The reason is that the strong coupling regime can be understood as the (vacuum) state where all the electrical fluxes are sharply vanishing. Consequently, the spin network basis can be seen as describing \emph{electric} flux excitations, that is non-vanishing electrical flux, with respect to this vacuum state. In loop quantum gravity, this state is referred to as the Ashtekar-Lewandowski vacuum \cite{Ashtekar1993, Ashtekar:1994mh} and represents a state of no spatial geometry.\footnote{That is operators measuring spatial areas and volumes have vanishing expectation value and vanishing fluctuations.} 

The previous remarks raise the question of whether there is a basis that describes \emph{curvature} (or \emph{magnetic}) excitations, generated from a vacuum state sharply peaked on flat connection. Such a basis would be adjusted to the weak coupling regime of lattice gauge theory and it would be very useful for the description of topological phases and their defect excitations \cite{Barkeshli:2014cna, Moradi:2014cfa,Hu:2015dga,Delcamp:2017pcw}. Defect magnetic excitatons on top of this alternative vacuum state are excitations leading to non-trivial values for Wilson loops based on contractible cycles. Such excitations are concentrated on point like defects in (2+1)d (e.g. the vertices of a 2d triangulation) and on one-dimensional defects in (3+1)d (e.g. the edges of a 3d triangulation). A basis in which the description of such defects is transparent would naturally help to understand their properties, such as their fusion and their braiding relations. As we describe further below, such a basis has been defined for (2+1)d models \cite{DDR1}. In this paper, we are interested in generalizing this basis to (3+1)d for which defect excitations are much less understood. 

Another reason to look for such a curvature basis are recent developments in loop quantum gravity.
The description of the Hilbert space of the theory has initially been based on the Ashtekar-Lewandowski vacuum (as a cyclic state of this Hilbert space), and the spin network basis has been very useful to describe the (electric flux) excitations on top of this vacuum.  Recently, new and inequivalent Hilbert space descriptions have been introduced, based on a vacuum state peaked on flat connections \cite{DGflux,DGfluxC,DGfluxQ, Lewandowski:2015xqa,Drobinski:2017kfm, Delcamp:2018sef} and vacua states peaked on homogeneously curved geometries \cite{DGTQFT,Dittrich:2017nmq}, respectively.  This new Hilbert space supports curvature (or magnetic) excitations localized on zero-dimensional (in two dimensions) or one-dimensional (in three dimensions) defect structures.   It is therefore useful to have a basis which includes labels for such curvature  excitations.

\bigskip \noindent
\textsc{Previous results:}
The approach we take in this paper is motivated by a number of recent results concerning the construction of bases of excited states. For (2+1)d lattice gauge theories, a basis which describes excitations on top of a flat connection vacuum can be obtained \cite{DDR1} by adapting the so-called \emph{fusion basis} from the theory of (2+1)d topological phases \cite{KKR, Hu:2015dga}. This fusion basis does not only describe magnetic excitations---which are encoded in conjugacy class labels $C$---but also electric excitations labeled by representations $R$ of the stabilizer groups of the conjugacy classes. Together, these labels form the irreducible representations $\rho=(C,R)$  of an algebraic structure known as the \emph{Drinfel'd double} of the group $G$. 

The fusion basis  provides labels for the basic excitations (e.g. those based on elementary plaquettes of the underlying lattice) but also for the excitations that arise from the fusion of the basic excitations. As such, this basis has an inherent hierarchical structure and is ideally suited for \emph{coarse-graining} \cite{Livine2013, DDR1}. In fact, it exhibits a crucial feature of non-abelian gauge theories, namely that the fusion of two purely magnetic excitation can lead to an excitation with dyonic charge, that is where both the magnetic and electric component are non-trivial \cite{deWildPropitius:1995hk}. This feature makes coarse-graining in the spin network basis quite cumbersome \cite{Livine:2016vhl,Delcamp:2016dqo, DGfluxC}. Another property of the fusion basis is that it is also valid when one replaces the gauge  group with a quantum group (for instance $\SU(2)_\k$) and, correspondingly, the fusion category of group representations with the fusion category of quantum group representations.  In this latter scenario, the fusion basis involves the so-called \emph{Drinfel'd centre} of this fusion category. 

Topological phases and their possible defects are much less understood in three spatial dimensions than in the lower dimensional case. As part of an ongoing attempt to fill this gap \cite{Wang:2014oya, 2012FrPhy...7..150W, wang2014braiding, Moradi:2014cfa, Wan:2014woa, Bullivant:2016clk, Wen:2016cij, Delcamp:2016lux, Dittrich:2017nmq, Williamson:2016evv, 2017arXiv170404221L, 2017arXiv170202148E, Riello:2017iti, Delcamp:2018wlb}, the authors proposed in \cite{Delcamp:2016lux} a procedure which lifts a (2+1)d topological phase to a (3+1)d one that is defined on a three-manifold ${\cal M}$, with possible excitations along the one-skeleton of a discretization $\triangle$ embedded in ${\cal M}$.  The basic idea is that the one-skeleton can be used to obtain a so-called \emph{Heegaard splitting} of the manifold ${\cal M}$.  This Heegaard splitting allows to encode the topology of ${\cal M}$ onto a Heegaard surface, which is equipped with two sets of curves, namely one-handle curves and two-handle curves. The Hilbert space for the (3+1)d theory can then be obtained by adopting the Hilbert space of a two-dimensional topological theory for the Heegaard surface. One has to furthermore implement constraints that prohibit degrees of freedom associated with the two-handle curves.  On the other hand, the one-handle curves are allowed to carry degrees of freedom and as such they describe excitations associated with the chosen one-skeleton.

This strategy was applied in \cite{Dittrich:2017nmq} to the Turaev-Viro model \cite{Turaev:1992hq} for $\SU(2)_\k$, which can be understood as a quantum deformation of three-dimensional $BF$ theory. This led to a consistent construction of a Hilbert space associated to the Crane-Yetter model---a quantum deformation of four-dimensional $BF$ theory.  But this (3+1)d Hilbert space does allow for curvature excitations along the one-skeleton of a discretization.  For this reason, one can identify this Hilbert space with the gauge invariant Hilbert space of a quantum deformed $\SU(2)_\k$ lattice Yang-Mills theory. Furthermore, this model displays a beautiful self-duality\footnote{In lattice gauge theory one exponentiates the connection to holonomies, which are then group valued, whereas the electric fluxes remain (Lie) algebra valued. Using a quantum group at root of unity one effectively also exponentiates the electric fluxes, which explains why such a self-duality might arise.}, see also \cite{Riello:2017iti}.  This is exhibited by the existence of two bases that are dual to each other: A (quantum deformed) spin network basis which diagonalizes certain invariant combinations of (quantum deformed) electrical flux operators and a \emph{dual spin network basis} that diagonalizes  Wilson loop operators. Both bases are labeled by the same algebraic data, namely the irreducible representations of $\SU(2)_k$. 

It turns out that both bases can be derived from two different choices of a (2+1)d fusion basis associated to the Heegaard surface. The definition of a fusion basis relies upon a so-called \emph{pant decomposition} of the corresponding surface. In the case of Heegaard surfaces, there are two natural choices for such a decomposition that are adjusted to the one-skeleton of the triangulation (i.e. the defect structure) or the one-skeleton of the dual complex, respectively.  The former yields the dual spin network basis, which encodes curvature excitations, while the latter yields the spin network basis.

These results for the quantum group $\text{SU}(2)_\k$ (which generalize to so-called modular fusion categories) open the question whether a similar dual spin network basis exist in the case of classical groups (which lead to non-modular fusion categories).  In this work, we use the same strategy involving Heegaard splittings to arrive at different parametrizations for the gauge invariant Hilbert space of (3+1)d lattice gauge theories.   We will see that we regain the spin network basis without any problems, but that we need to provide in general more input to construct a complete dual spin network basis.  Therefore, the case of classical groups turns out to be more complicated than for quantum groups.\footnote{ This can be understood from the following fact: For the (2+1)d theories the algebraic structure of the fusion basis is determined by the Drinfel'd double of the representation category of the group or quantum group. In both cases this defines a modular fusion category.  
For the (3+1)d case we have to impose the two-handle constraints onto the fusion basis states. As we mentioned above, if the fusion basis is adapted to the dual graph, this results in the spin network basis. That is the algebraic structure is reduced from the Drinfel'd double to the original fusion category of representations of the group or quantum group, which are non-modular and modular, respectively.}

We wish here to concentrate on the algebraic aspects and  avoid rather involved measure theoretic issues.  We thus consider only finite groups. Finite group models can be also used to test coarse-graining algorithms for spin foams and lattice gauge theories \cite{Bahr:2011yc,Dittrich:2014mxa,Delcamp:2016dqo}. We follow the strategy of using the Heegaard surface representation of three-dimensional manifolds to obtain convenient representations of the gauge invariant subspace of three-dimensional graph connections. To this end we need, to understand the parametrization  of the space of flat connection on the Heegaard surface. This allows us to isolate the two-handle constraint equations that will determine the structure of the dual spin network basis. We also show that the strategy of using Heegaard splittings is useful in order to describe the Hilbert space of connections associated to three-dimensional manifolds with boundary. We will see that there are different ways of defining such a boundary, which can be seen to be adjusted to either the spin network basis or the dual spin network basis. The latter gives a more direct description of the excitations. By introducing a notion of cutting and gluing of three-dimensional manifolds and their associated state spaces, we can generalize Ocneanu's tube algebra in order to reveal the algebraic structure of the excitations associated with certain 3d manifolds.

\bigskip \noindent
\textsc{Plan of the paper:} In sec.~\ref{sec:simple_ex}, we illustrate the key mechanism of lifting the fusion basis from two to three dimensions using the simplest possible example, namely a curvature defect along a loop embedded into a three-sphere. This leads to (a simple version of) the spin network basis and the dual spin network basis, and display how these are connected via a so-called $S$-transform. We encounter the Drinfel'd double of the group and its representations, which we shortly review. We discuss two different parametrizations of the space of locally flat connections on a closed surface in sec.~\ref{sec:param}. This understanding of the space of locally flat connections for a surface will be crucial for the lifting procedure and the imposition of the constraints. We review this lifting procedure in sec.~\ref{sec:lifting} and then consider the lifting of the fusion basis, firstly adapted to the dual graph, and secondly adapted to the one-skeleton of the triangulation.  This first case yields the spin network basis.  The second case yields a basis describing directly the curvature or magnetic excitations. But we will see that in general we need to specify further observables to obtain a complete basis. We consider explicit examples and  determine these additional observables and the label sets that allow to obtain a complete dual spin network basis. 
We subsequently consider three-manifolds with boundary in sec.~\ref{sec:cutglue}. We will see that the Heegaard surface splitting allows in particular for a description of boundaries slicing through the faces and edges of the triangulation (as opposed to the faces and edges of the dual complex). This description allows us to derive a generalization of the so-called quantum triple algebra (itself a generalization of the tube algebra).

\section{A simple example \label{sec:simple_ex}}
\subsection{Three-sphere with a loop defect}

We are interested in describing the space of gauge invariant states of a lattice gauge theory defined on the one-skeleton  $\Upsilon_1$ of a three-dimensional cell complex  $\Upsilon$ that is embedded into a three-dimensional manifold $\mathcal{M}$.  We declare locally flat configurations, that is configurations whose holonomies along all the  loops contractible in $\mathcal{M}$ are trivial, as vacuum states.  Excited states are thus states which have a non-trivial holonomy for at least one contractible curve. We can however render a set of closed curves in $\mathcal{M}$ non-contractible by removing an appropriate one-dimensional defect structure. For instance, if we want to produce a new non-contractible loop, we remove the blow-up of a loop, which encircles the first loop, from ${\cal M}$. Generalizing this procedure, we can identify the gauge invariant state space of connections on $\Upsilon_1$  with the space of gauge invariant flat connection on $\mathcal{M} \backslash \blow$ where $\blow$ is the blow-up of the one-skeleton of the complex $\triangle$ dual to $\Upsilon$.  

In this section, we explain the main concepts we will use in the rest of this work using a simple example: 
 We choose the three-manifold $\mathcal{M}$  to be the the three-sphere $\mathbb{S}_3$ and consider a closed loop embedded in $\mathbb{S}_3$ as defect structure. By considering a regular neighborhood (or less precisely a blow-up) of this closed loop, we obtain a solid two-torus denoted by $\mathring{\mathbb{T}}_2$. It turns out that the manifold $\mathbb{S}_3 \backslash \mathring{\mathbb{T}}_2$ is also isomorphic to a solid two-torus. This defines a decomposition of the three-sphere as $\mathbb{S}_3 = \mathring{\mathbb{T}}_2 \cup \mathbb{S}_3 \backslash \mathring{\mathbb{T}}_2 \simeq \mathring{\mathbb{T}}_2 \cup \mathring{\mathbb{T}}_2$ which states that the three-sphere can be obtained as the gluing of two solid tori. This decomposition is referred to as the \emph{genus one Heegaard splitting} of the three-sphere. The gluing of the two tori is performed along the boundary $\partial \mathring{\mathbb{T}}_2$ of the manifold $\mathring{\mathbb{T}}_2$ which is by definition homeomorphic to the two-torus $\mathbb{T}_2$, and defines the Heegaard surface associated to the Heegaard splitting above.
 
Notice that any flat connection in $\mathbb{S}_3 \backslash \mathring{\mathbb{T}}_2$ can be mapped to a flat connection on its boundary $\mathbb{T}_2$. But the space of flat connections on the torus is larger than the space of flat connections on $\mathbb{S}_3 \backslash \mathring{\mathbb{T}}_2$. This is due to a set of curves, generated by a loop parallel to the defect loop, which are non-contractible on the torus but are contractible in $\mathbb{S}_3 \backslash \mathring{\mathbb{T}}_2$. We can thus define the state space of flat connections in  $\mathbb{S}_3 \backslash \mathring{\mathbb{T}}_2$ by following two steps:
$(i)$ Define the state space of flat connections on the surface $\mathbb{T}_2$. $(ii)$ Impose that the holonomies along the contractible cycles in $\mathbb{S}_3 \backslash \mathring{\mathbb{T}}_2$ are  flat.

\begin{figure}[t]
	\centering
	\label{fig:example}
	\includegraphics[scale=0.85,valign=c]{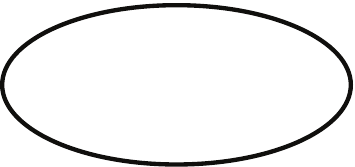} $\longrightarrow$ 
	\includegraphics[scale=0.85,valign=c]{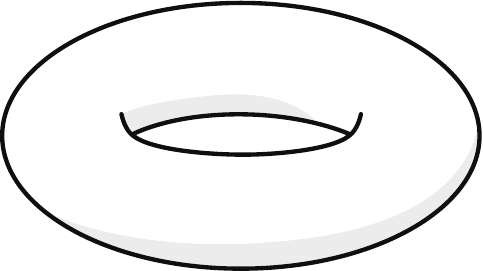} $\longrightarrow$
	\includegraphics[scale=0.85,valign=c]{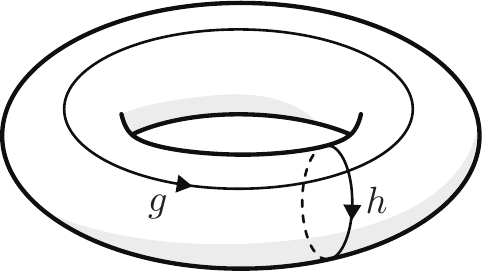}
	\caption{The left panel represents a closed line embedded into the three-sphere. The middle panel corresponds to a regular neighborhood of such closed line which is nothing else than a solid two-torus $\mathring{\mathbb{T}}_2$. The right panel represents a graph embedded on the boundary of the solid two-torus which captures the two non-contractible cycles.}
\end{figure}

Let us therefore consider the state space of flat connections on the two-torus. The two-torus has two non-contractible cycles $\ell_1$ and $\ell_2$ which satisfy the relation $\ell_1\circ \ell_2 \circ\ell_1^{-1}\circ \ell_2^{-1} \simeq \mathrm{triv}$, where $\ell=\mathrm{triv}$ denotes the trivial cycle. These cycles, which are represented in fig.~\ref{fig:example} are referred to as the \emph{meridional} and the \emph{equatorial} cycle, respectively. Let us denote by $|g,h \ra_{\tor}$ the graph-states defined on $\mathbb{T}_2$ where the group variables $g,h \in G$ label the equatorial and the meridional cycle, respectively. The Hilbert space $\mathcal{H}_{\mathbb{T}_2}$ of gauge invariant functionals on the space of flat connections on $\mathbb{T}_2$ is then spanned by states\footnote{An inner product on this space can be  induced from an inner product $\la g,h | g',h'\ra  =\delta_{g,g'}\delta_{h,h'}$ on $L^2(G\times G)$.}  
\begin{equation}
	\label{gsTorus}
	\mathcal{H}_{\mathbb{T}_2} = 
	\big\{ 
	\frac{1}{\sqrt{|G||Z_{g,h}|}}\sum_{x \in G}|xgx^{-1},xhx^{-1}\ra \, | \, [g,h]=\unitG
	\big\}
\end{equation}
where $[g,h] := ghg^{-1}h^{-1}$ is the group commutator and $Z_{g,h} = \{x \in G \, | \, [x,h] = \unitG = [x,g]\}$ is the stabilizer group of the tuple $(g,h)$ under the adjoint action.  The averaging over $x \in G$ is there to ensure gauge invariance at the single vertex.
The dimension of this Hilbert space is given by
\be
	{\rm{dim}}(\mathcal{H}_{\mathbb{T}_2} )= \big|\{(g,h)\in G^2 \, | \,	 [g,h]=\mathbbm{1}_G\}/G \big| 
\ee 
where it is understood that the group $G$ acts by conjugation. 

To obtain the Hilbert space of flat connections on $\mathbb{S}_3 \backslash \mathring{\mathbb{T}}_2$ we have to impose that cycles, which are not contractible in  $\mathbb{T}_2$ but are contractible in $\mathbb{S}_3 \backslash \mathring{\mathbb{T}}_2$, carry a trivial holonomy. Such cycles are generated by the equatorial cycle labeled by the $g$-holonomy. Thus the Hilbert space is spanned by states
\be\label{Ctorus states}
	|C \ra_{\tor}  :=\frac{1}{\sqrt{|C|}}\sum_{h \in C}| \unitG, h \ra_{\tor} 
\ee
where $C$ denotes a conjugacy class of $G$.
We would have also obtained this basis of states if we had started with a particular\footnote{We will soon encounter another fusion basis.} \emph{fusion basis} for the state space associated to the torus $\mathbb{T}_2$.  This basis, which we will present in detail further, is parametrized by a conjugacy class $C$ of $G$ and an irreducible representation $R$ of the stabilizer group $Z_C$ of $C$. This stabilizer group  is defined as $\{g \in G \, | \, [g,c_1]=\unitG\}$ with $c_1$ being the representative of $C$.  The conjugacy class $C$ characterizes the holonomy $h$ of the meridional cycle and the representation $R$ characterizes the dependence of the state on the equatorial holonomy $g$. The alternative basis states are then given by
\be
\label{gsTorusF1}
	| R,C \ra_{\mathbb{T}_2} = \frac{1}{\sqrt{|G|}}\sum_{q \in G/Z_C \atop z \in Z_C}\chi^{R}(z) \; | q z q^{-1}  ,  q c_1 q^{-1} \ra_{\mathbb{T}_2} 
\ee
where $\chi^R$ denotes the character of the representation $R$ of $Z_C$. The states \eqref{Ctorus states} can be written as the following linear combinations of these fusion basis states 
\ba 
|C \ra_{\tor} &=&\frac{1}{\sqrt{|C|}}\sum_{h \in C}|\unitG,h \ra_{\tor} \nn\\
&=&
\frac{1}{|Z_C| \sqrt{|C|}}\sum_{q \in G/Z_C \atop z \in Z_C}\sum_R
	d_R\chi^R(z) \; |qzq^{-1},qc_1q^{-1} \ra_{\mathbb{T}_2} \\
&=&\frac{1}{\sqrt{|Z_C|}}	\sum_R d_R | R,C \ra_{\tor} 
\ea
where we made use of the well-known identity for finite groups
\be
\frac{1}{|Z_C|}\sum_R d_R \chi^R(z) = \delta_{z,\mathbbm{1}_{Z_C}} \; .
\ee

\begin{figure}[t]
	\centering
	\label{fig_tori}
	\includegraphics[scale=0.85,valign=c]{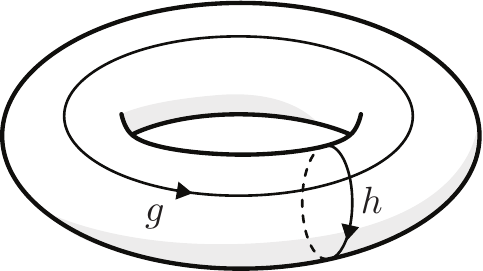} \q $\stackrel{S}{\longleftrightarrow}$ \q
	\includegraphics[scale=0.85,valign=c]{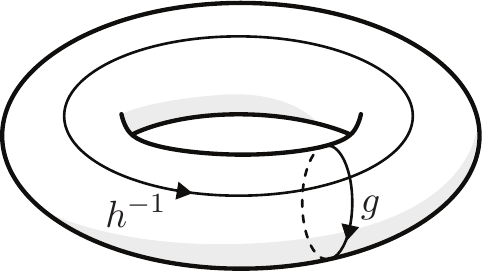}
	\caption{Graphical depiction of the $S$-transform which sends the graph-state $|g,h \ra$ to $|h^{-1},g \ra$. The matrix elements of this map relates the two fusion bases which can be defined on the two-torus.}
\end{figure}

\noindent
We could have also started with a fusion basis where the conjugacy class $C$ characterizes the equatorial holonomy and $R$ the meridional one. These basis states are given by
\begin{align}
	\label{gsTorusF2}
	|C,R \ra_{\tor} &= \frac{1}{\sqrt{|G|}}\sum_{q \in G/Z_C \atop z \in Z_C}\chi^{R}(z) \,|qc_1^{-1}q^{-1},qzq^{-1} \ra_{\tor} 
\end{align}
and are related to the first set of fusion basis states \eqref{gsTorusF1} via a so called \emph{$S$-transformation}. This transformation has a simple geometric interpretation: It is a large diffeomorphism of the torus that exchanges the equatorial with the meridional cycle, see fig. \ref{fig_tori}.

Using this second set of fusion basis states, it is straightforward to implement the triviality of the equatorial holonomy,  as it just requires to set $C$ to be the trivial conjugacy class:
\begin{equation}
	|R \ra_{\tor} = \frac{1}{\sqrt{|G|}}\sum_{h \in G}\chi^{R}(h)|\unitG, h \ra_{\tor} \; . 
\end{equation}
Here the label $R$ stands for an irreducible representation of the group $G$ (which is the stabilizer of the trivial conjugacy class). Thus, starting from the two possible fusion bases for the torus, we have obtained two different bases for the space of flat connections on $\mathbb{S}_3 \backslash \mathring{\mathbb{T}}_2$. The first basis, which is labeled by a conjugacy class does characterize the (magnetic or curvature) excitation along the defect loop. The second basis is labeled by a representation of the group and can be identified with a spin network basis based on a loop dual to the defect loop. 
The two bases can be transformed into each other by a restriction of the $S$-transform to the subspace of flat connections for which the equatorial connection is trivial.  In terms of the 3d theory, this leads to a gauge invariant version of the group Fourier transform: 
\be\label{trafoSNWC}
	|C \ra_{\tor} \,=\,\sum_R  \frac{1}{\sqrt{|Z_C|}}\overline{\chi^R(C)}|R\ra_{\tor} \q , \q |R \ra_{\tor} \,=\, \sum_{C}\frac{1}{\sqrt{|Z_C|}} \chi^R(C) \,|C\ra_{\tor} \; .
\ee

\subsection{Tube algebra and Drinfel'd double \label{sec:tube}}

In the previous section, we defined basis states for the Hilbert space of gauge invariant functionals on the space of flat connections on the torus. These basis states were labeled by a pair $(C,R)$ with $C$ a conjugacy class and $R$ an irreducible representation of the stabilizer $Z_C$. In this section, we review  how this pair labels the irreducible representations of a well-known algebraic structure, namely the Drinfel'd double of the gauge group $G$ \cite{Drinfeld:1989st, Koornwinder1999}. This algebraic structure arises by considering the gluing of states defined on cylinders \cite{ocneanu1993, ocneanu2001, Lan2013, DDR1, Delcamp:2017pcw}. The representation theory of the Drinfel'd double can then be used to define  the \emph{fusion basis} on arbitrary (handle-body) surfaces.

\begin{figure}[t]
	\centering
	\includegraphics[scale=0.85,valign=c]{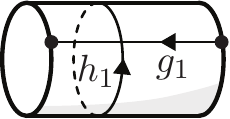} $\star$ 
	\includegraphics[scale=0.85,valign=c]{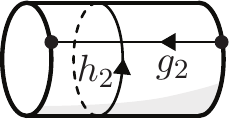} =
	$\mathbb{P} \; \triangleright$ 
	\includegraphics[scale=0.85,valign=c]{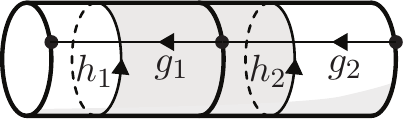}
	$\sim$ $\delta_{h_1,g_1h_2g_1^{-1}}$
	\includegraphics[scale=0.85,valign=c]{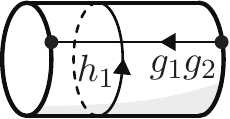}
	\caption{Graphical depiction of Ocneanu's tube algebra. The cylinders are glued by identifying the punctures and the links ending at these punctures are connected. This leads to a new internal two-valent node and a new face (represented in gray on the figure) at which we need to enforce gauge invariance and flatness, respectively. The resulting state is gauge invariant and flat and can therefore be expressed on a minimal graph.  The result defines the multiplication rule of the Drinfel'd double.}
	\label{fig_algebra}	
\end{figure}

The cylinder surface denoted by $\cyl$ can be obtained from a two-sphere by removing two disks from the surface. In addition, the boundary of each of these disks is required to carry a single \emph{marked point} (see fig.~\ref{fig_algebra}).
We define a graph on the cylinder, seen  in fig.~\ref{fig_algebra}, 
which captures the single non-contractible cycle and such that there is an open link ending at  each of the marked points.   We consider the space of connections for this  graph and impose gauge invariance at the four-valent node but do not require gauge invariance at the one-valent nodes. 
By using a gauge fixing, we can define a basis for such connection states  $|g,h\ra_{\cyl}$  where $(g,h)\in G\times G$ denote the holonomies as represented in fig.~\ref{fig_algebra}.  A general state is then  given by 
\ba
\psi \,=\, \sum_{g,h} \psi(g,h) |g,h \ra_{\cyl} \; .
\ea
Two cylinders can be glued to each other such that they give another cylinder. We will now define a gluing of the corresponding state spaces such that the resulting states define again (flat) connection states on the cylinder. This turns the space of connections on the cylinder into an algebra.

Let us define the general gluing procedure (see \cite{Delcamp:2016eya, Delcamp:2017pcw} for more details). Let $\Sigma_A$ and $\Sigma_B$ be two surfaces, $\Gamma_A$ and $\Gamma_B$ two graphs embedded on $\Sigma_A$ and $\Sigma_B$, respectively, and $\mathcal{H}_A$ and $\mathcal{H}_B$ the corresponding gauge invariant Hilbert spaces of locally flat connections. Both $\Sigma_A$ and $\Sigma_B$ are assumed to have boundaries whose components carry single marked points. The embedded graphs end at these boundaries by at least one open edge. The gluing of the surfaces $\Sigma_A$ and $\Sigma_B$ is a surface $\Sigma_{A \cup B}$ obtained by identifying the boundary of the disks defining the punctures from each one of them, as well as the corresponding marked points. Furthermore, the links of the embedded graphs along which the gluing is performed must be connected. The result is a single embedded graph $\Gamma_{A \cup B}$. This resulting graph  possesses new internal nodes and new closed faces. A connection state for the graph $\Gamma_{A \cup B}$ might therefore not be gauge invariant at these new internal nodes and also not be flat at the new internal faces. Therefore, we introduce a projector $\mathbb{P}$ which performs a gauge averaging at the internal nodes and multiplies the states with a delta function $\delta(h_f,\unitG)$ for each new closed face $f$. The resulting Hilbert space is denoted $\mathcal{H}_{A \cup B}$.  Note that states in this Hilbert space can be mapped bijectively to states on a minimal graph for $\Sigma_{A\cup B}$.

The gluing procedure $\star$ between states $\psi_A \in \mathcal{H}_A$ and $\phi_B \in \mathcal{H}_B$ is  defined  to be the projection   into $\mathcal{H}_{A \cup B}$ of the product of the two states:
\begin{align}
	\label{starProd}
	\star \; : \;  \mathcal{H}_A \times \mathcal{H}_B 
	&\longrightarrow \mathcal{H}_{A \cup B} \\ \nn
	(\psi_A, \phi_B) &\longmapsto \mathbb{P} \triangleright (\psi_A \cdot \phi_B)
\end{align}
Let us now apply this procedure to the gluing of two cylinders. In this case, the map $\star$ defines an algebra product since the gluing of two cylinders leads to another cylinder. This algebra is known as Ocneanu's tube algebra \cite{ocneanu1993, ocneanu2001, Lan2013, DDR1, Delcamp:2017pcw}. We consider two basis states $|g_1,h_1\ra_{\cyl}$ and $|g_2,h_2\ra_{\cyl}$ living on two different cylinders. The gluing is graphically depicted fig.~\ref{fig_algebra} and the result reads
\be \label{DDproduct}
	|g_1,h_1 \ra_{\cyl} \star | g_2 ,h_2 \ra_{\cyl} =  \delta_{h_1,g_1 h_2 g_1^{-1}}|g_1 g_2,h_1\ra_{\cyl} \; .
\ee
This fundamental result shows that the gluing of two cylinders reproduces the multiplication rule of the Drinfel'd double $\mathcal{D}(G)$ whose definition is recalled in app.~\ref{app_drinfeld}. 

Once we have identified the algebra, we can ask for the corresponding irreducible linear representations. Physically, the irreducible modules are characterized by observables (identified as charges) that are left unchanged under the gluing. In other words, we can think of the multiplication $|g_1,h_1 \ra_{\cyl} \star |g_2,h_2 \ra_{\cyl}$ as the action of the algebra element $g_1 \smo \delta_{h_1}$ on the state $| g_2, h_2 \ra_{\cyl}$. We are then asking for observables, which are identified with charges, that are invariant under this action. From the algebra \eqref{DDproduct} we can indeed see that the conjugacy class $C$ of the $h$-holonomies is preserved. We furthermore see that the algebra reproduces for the $g$-holonomies the usual group product. We can therefore expect a second invariant observable involving irreducible representations as functions of the $g$-argument.  However, we also have to take into account that the $g$-holonomies act on the $h$-holonomies. Indeed, the irreducible representations  of the Drinfel'd double $\rho=(C,R)$ are labeled by a conjugacy class $C$ of $G$ and a representation $R$ of its stabilizer group $Z_C$. The matrix elements of these representations can be used to define a new basis for the cylinder states given by (see app.~\ref{app_drinfeld} for details): 
\be
	\label{cyl}
	|\rho,MN \ra_{\cyl} = \frac{1}{\sqrt{|G|}}\sum_{g,h \in G}\sqrt{d_{\rho}}D^{\rho}_{MN}(g \smo \delta_h) \; |g,h \ra_{\cyl} \; .
\ee  
Here $D^{\rho}_{MN}$ denotes a matrix element of an irreducible representation $\rho$ of $\mathcal{D}(G)$ and $d_\rho$ its dimensions. We denote by $g\smo \delta_h$ a basis element of the Drinfel'd double algebra. 
We will refer to the  states $| \rho, MN \ra_{\cyl}$ as the \emph{fusion basis states} for the cylinder.

By construction the new basis states diagonalize the $\star$-product:
\be\label{gr2}
	|\rho_1, M_1N_1 \ra_{\cyl} \star | \rho_2 ,M_2N_2 \ra_{\cyl}
	 = \sqrt{|G|}\frac{\delta_{\rho_1,\rho_2}}{\sqrt{d_{\rho_1}}}\delta_{N_1,M_2}
	 | \rho_1, M_1N_2 \ra_{\cyl}  \; ,
\ee
making explicit that the charges, which are characterized by the representation label $\rho=(C,R)$, are invariant under the gluing. The conjugacy class $C$ describes the magnetic excitation, that is the trace of the holonomy around the cylinder. The representation $R$ (of the stabilizer group of $C$) characterizes the electric excitation, in the sense that it describes the dependence of the state on the holonomy along the cylinder. When the representation $R$ is trivial, the state is completely gauge invariant and has therefore a vanishing electrical charge.

Note that the gluing rule \eqref{gr2} can be used to regain the states for the torus, namely by gluing cylinder states. 
The torus is obtained  by identifying the two punctures and the corresponding marked points of a cylinder. After identification, the two one-valent nodes become a two-valent vertex at which the group averaging must be applied. This induces a contraction of the corresponding representation space indices. The flatness is also implemented, as the character has only support on elements $g\smo \delta_h$ for which $g$ and $h$ commute.  The resulting state is given by 
\be
	\label{torus}
	| \rho \ra^A_{\tor} = \frac{1}{\sqrt{|G|}}\sum_{g,h}\chi^{\rho}(g \smo \delta_h)|g,h \ra_{\mathbb{T}_2} 
\ee
where the explicit expression for the characters $\chi^\rho$ of the Drinfel'd double can be found in app.~\ref{app_drinfeld}. Using the explicit formula for the characters, this recovers the states \eqref{gsTorusF1} defined previously.

As a matter of fact, the torus can be obtained from the gluing of a cylinder in two different ways---the gluing can be performed along the equatorial or the meridional cycle. Correspondingly, there is a second fusion basis for the torus which is related by the so-called $S$-transform
\ba
S: \, g \smo \delta_h \,\, \mapsto\,\, h^{-1} \smo \delta_g
\ea
to the first one. The states of this second fusion basis are given by 
\be	
	| \rho \ra^B_{\tor} = \frac{1}{\sqrt{|G|}}\sum_{g,h}\chi^{\rho}(h \smo \delta_{g^{-1}})|g,h \ra_{\mathbb{T}_2} \; .
\ee
This can be used to compute the matrix elements of the $S$-transform as the transformation which relates these two bases (see app.~\ref{app_puncS}).  Applying this transformation to states that satisfy the additional flatness constraint $g=\mathbbm{1}_G$ we recover the transformation rules \eqref{trafoSNWC} between the curvature excitation basis and the spin network basis for the flat connection states on $\mathbb{S}_3 \backslash \mathring{\mathbb{T}}_2$.

\section{Parametrizations for the  space of  flat connections on 2d surfaces \label{sec:param}}
In the previous section, we illustrated with a simple example our strategy to define Hilbert spaces and bases for the space of flat connections on three-dimensional manifolds with defect structures. This strategy can be summarized as follows: Firstly, we define the Hilbert space of gauge invariant functions on the space of flat connections on a two-dimensional surface obtained from a Heegaard splitting, which is adjusted to the defect structure.  Secondly, we impose that the holonomies along the  cycles that are contractible in the three-dimensional manifold with defect structre are trivial. In this section, we focus on the first one of these steps, namely the construction of the Hilbert space of flat connection on two-dimensional surfaces. We will describe two different holonomy parametrizations. The first one gives a global description, whereas the second one, based on a pant decomposition of the surface, is more local. This second description leads to the fusion basis and this is also the one we use to impose the flatness constraints needed for the construction of the 3d state spaces.

\subsection{Holonomy parametrization} \label{holp}

\begin{figure}[t]
	\centering
	\includegraphics[scale=0.85,valign=c]{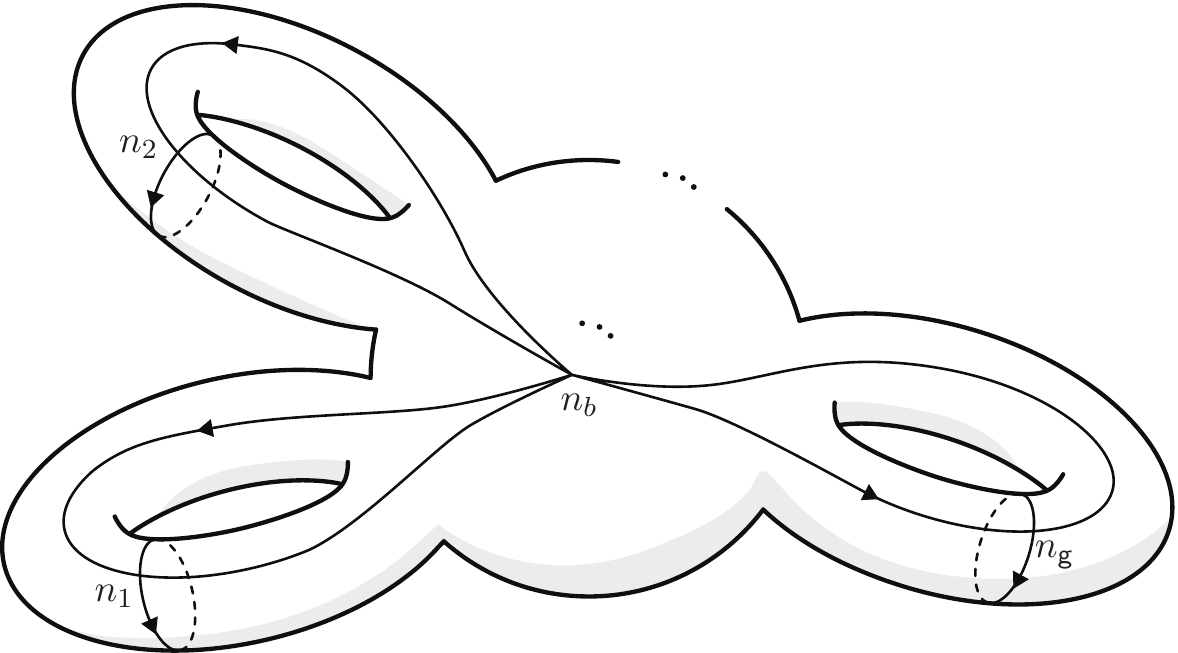}  
	\caption{A genus-$\mathsf{g}$ two-dimensional hypersurface can be obtained by gluing $\mathsf{g}$ cylinders to a $2\mathsf{g}$-punctured two-sphere. We then define a graph $\Gamma$ on the surface which captures all the non-contractible cycles. }
	\label{fig_handles}
\end{figure}

The first parametrization we discuss is based on a minimal set of holonomies. Let $\Sigma$ be a genus-$\mathsf{g}$ two-dimensional surface. It is possible to represent $\Sigma$ as a sphere with $\mathsf{g}$ handles glued to it (see fig.~\ref{fig_handles}). More precisely, we can obtain $\Sigma$ by gluing twice-punctured two-spheres (or cylinders) to a $2\mathsf{g}$-punctured two-sphere. Fixing a base point $n_b$ on the sphere, we choose for each handle $i$ an oriented curve that starts and ends at $n_b$ and by going along the handle $i$ only. The orientation of the curve induces an orientation for the handle which allows to differentiate between the \emph{source} and the \emph{target} punctures on the sphere to which the handle is glued. Furthermore, to every curve going along a handle, we assign a node $n_i$ that is located on the curve, as well as another curve starting and ending at $n_i$ that goes around this (and only this) handle once.  The orientation of this curve is chosen such that it goes anti-clockwise around the corresponding source puncture as seen from the target puncture. The resulting graph is denoted by $\Gamma$.

Let us now define a \emph{graph connection} on this graph $\Gamma$ by assigning $g_i$-holonomies to the links going from the base point $n_b$ to the node $n_i$ on the handle $i$ and $h_i$-holonomies to the links going around the handle $i$ from $n_i$ to $n_i$.  The remaining links, i.e. the links going from the nodes $n_i$ to the base point $n_b$, are labeled by a trivial holonomy. This can be understood as a gauge fixing condition for the gauge action at the nodes $n_i$.

So far we have defined a graph connection on $\Gamma$ by assigning a set of group elements $\{g_i,h_i\}_{i=1}^{\mathsf{g}}$. But this connection is not necessarily flat. In order to enforce the flatness constraint, we need to impose that contractible cycles are associated with trivial holonomies. Let us consider the path going around every puncture on the sphere. Such a path can be contracted to a trivial path and therefore the corresponding holonomy must be trivial. This flatness condition can be also understood as imposing the Bianchi identity for the sphere. To give the flatness constraint explicitly we assume  that the links from the base point to the handles $i$ can be cyclically ordered around $n_b$, without any crossings, as follows: If we denote by $\ell_i$ the link from $n_b$ to $n_i$ and by $\ell_{i}'$ the link from $n_i$ to $n_b$ the cyclic ordering in the clockwise direction is given by $(\ell_1,\ell_1',\ell_2,\ell_2',\dots,\ell_{\mathsf{g}},\ell_{\mathsf{g}}')$. The flatness constraint finally reads
\begin{equation}
	\label{Flatness1}
	 \prod_{i=1}^{\mathsf{g}}[g_i,h_i] \,=\, \unitG \; .
\end{equation}
Furthermore, there is a remaining gauge action at the base point $n_b$. This leads to an adjoint action on all group elements $\{g_i\}$ and $\{h_i\}$ since we assume that the gauge fixing discussed above remains intact:
\ba\label{Gauge1}
	\{g_i,h_i\}_{i=1}^{\mathsf{g}} \longrightarrow  \{G g_iG^{-1}, G h_i G^{-1}\}_{i=1}^{\mathsf{g}} \; .
\ea
We notice in particular that such simultaneous action by conjugation preserves the flatness constraint \eqref{Flatness1}. In summary, the space of flat connections on a genus-$\mathsf{g}$ surface is parametrized by equivalence classes 
\ba\label{Param1}
(G^{2\mathsf{g}})_{|\text{flat}} / \text{Ad}(G)
\ea
such that the flatness constraint \eqref{Flatness1} is satisfied. 

An inner product can be straightforwardly defined, e.g. by introducing a Hilbert space spanned by states $|\{g_i,h_i\}_{i=1}^{\mathsf{g}}\rangle$, and by defining 
\ba
\langle \{g_i,h_i\}_{i=1}^{\mathsf{g}} \,| \{g'_i,h'_i\}_{i=1}^{\mathsf{g}}\rangle &=& \prod_{i=1}^{\mathsf{g}} \delta(g_i,g'_i)\,\delta(h_i,h'_i) \; .
\ea
Notice that this inner product is invariant under the adjoint action \eqref{Gauge1}. Therefore, one can define an \emph{induced} inner product on the subspace of wave functions satisfying the constraint \eqref{Flatness1} and invariant under the action \eqref{Gauge1}. A similar strategy can be applied for the other holonomy parametrizations which we present below. In this case, one has a larger number of holonomy parameters to begin with and correspondingly has to implement a larger number of gauge invariances and flatness constraints. For this reason, the induced inner products obtained from different holonomy parametrizations might differ by overall factors of $|G|$.

\begin{figure}[t]
	\centering
	\includegraphics[scale=0.85,valign=c]{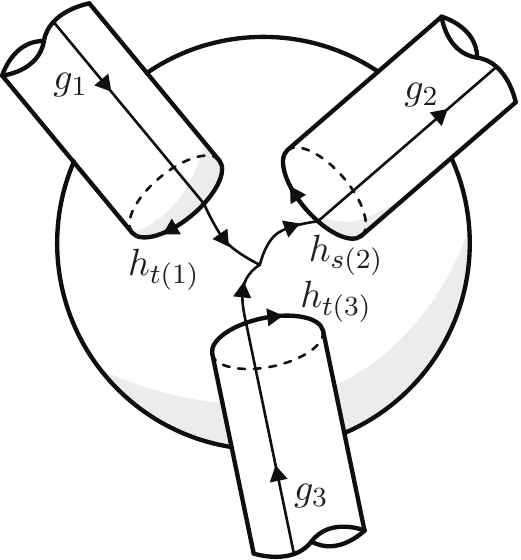}\q \q \q 
	\includegraphics[scale=0.85,valign=c]{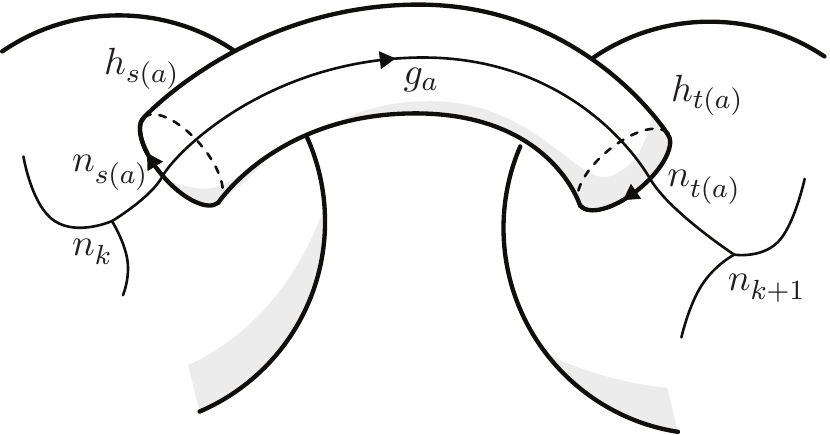}
	\caption{Graphical representation of the conventions used to label the holonomies and the nodes of the graph.}
	\label{fig_3puncsphere}
\end{figure}

\bigskip \noindent
Let us now present a more local holonomy parameterization for which we employ a more refined graph.  This more local holonomy representation will lead us eventually to the Drinfel'd Double parametrization. To define the more refined graph, we use the fact that every two-dimensional Riemann surface can be obtained as a gluing of thrice-punctured two-spheres denoted by $\mathbb{Y}$. More precisely, we can decompose a closed surface of genus $\mathsf{g}$ with $\mathsf{g} \geq 2$ into $2\mathsf{g}-2 $  thrice-punctured two-spheres $\mathbb{Y}$.\footnote{This simply follows from Euler's formula which states that for a convex three-dimensional polyheron: $\#_{\rm loops} = \#_{\rm edges} - \#_{\rm vertices} + 1$. Since we are looking for a decomposition into thrice-puncture spheres, we have the additional constraint that $3\#_{\rm vertices} = 2 \#_{\rm edges}$. Setting $\#_{\rm loops} = \mathsf{g}$, we finally obtain $\#_{\mathbb{Y}} = \#_{\rm vertices} = 2\mathsf{g}-2$.} We label the thrice-punctured spheres by $k=1, \ldots, (2\mathsf{g}-2)$ and we choose a base node $n^{\mathbb{Y}}_k$ on each sphere. Furthermore, as before, we assign the $(3\mathsf{g}-3)$ cylinders, which are glued to the punctures, with an orientation so that we can define source and target punctures associated to a given cylinder.  Source and target punctures are all equiped with a marked point living at the boundary of the corresponding disks. These marked points, which serve as nodes for the graph $\Gamma$, are denoted by $n^{\cyl}_{s(a)}$ and $n^{\cyl}_{t(a)}$ with $a=1, \dots, (3\mathsf{g}-3)$.  Putting everything together, we  construct the graph\footnote{We do not allow any crossing of the links accept at the nodes.} $\Gamma$ together with a graph connection by choosing:
\begin{enumerate}[itemsep=0.4em,parsep=1pt,leftmargin=*]
	\item[$\circ$] For each of the source $s(a)$ and target punctures $t(b)$  on a given sphere $k$, a link from the base point $n^{\mathbb{Y}}_k$ to the marked points $n^{\cyl}_{s(a)}$ and $n^{\cyl}_{t(b)}$, respectively. We then associate a  trivial holonomy to these links, or equivalently we use the gauge freedom at the marked points in order to gauge fix these to the identity.
	\item[$\circ$] For each cylinder $a$,  a link from its source node $n^{\cyl}_{s(a)}$ to its target node $n^{\cyl}_{t(a)}$. We associate a holonomy $g_a=g_{t(a)s(a)}$ to this link and such that the inverse holonomy $g_a^{-1}$ is denoted by $g_{s(a) t(a)}$.  
	\item[$\circ$] For each cylinder $a$, we define a link around the corresponding target puncture with {clockwise} orientation (as seen on the target sphere) which starts and ends at $n^{\cyl}_{t(a)}$. The corresponding holonomy is denoted by $h_{t(a)}$. Sometimes it is also convenient to consider links around the source punctures with clockwise orientation as seen on the source sphere. Flatness along the cylinder then imposes that the corresponding holonomy reads $h_{s(a)}:= g^{-1}_{a} h_{t(a)}^{-1}  g_a $ for these links.
\end{enumerate}
The graph associated with a thrice-punctured two-sphere following these conventions is represented in fig.~\ref{fig_3puncsphere}.  

As before, such a parametrization $\{g_a,h_a\}_{a=1}^{3\mathsf{g}-3}$ is over-complete since additional constraints and equivalence relations need to be imposed. Firstly, for each sphere there is a flatness constraint, which can be interpreted as the Bianchi identity for this sphere.  Consider a sphere $k$  and a clockwise ordering of the links around the base node $n^\mathbb{Y}_k$  given by $( (a_1,o_1), (a_2,o_2), (a_3,o_3))$ where $o=s,t$ denotes whether the puncture is a source or target one, respectively.  The flatness constraint is then given by 
\ba\label{Flatness2}
	h_{o_3(a_3)}  h_{o_2(a_2)}  h_{o_1(a_1)} =\unitG  \; .
\ea
Such expression typically involves $g_a$ holonomies via the definition  $h_{s(a)}:= g^{-1}_{a} h^{-1}_{t(a)}  g_a$. Moreover, there might be redundancies among the set of flatness constraints, e.g. for a genus three surface one finds only three independent flatness constraints. In general, we find that the number of independent flatness constraints matches the genus of the surface. Finally, we are looking for equivalence classes under the  gauge action at the base node $n^{\mathbb{Y}}_k$ of every thrice-punctured two-sphere. A transformation with gauge parameters $\{G_k\}_{k=1}^{2\mathsf{g}-2}$ acts as 
\begin{equation}
	\label{Gauge2}
	\{ g_{t(a) s(a)}, h_{t(a)} \}_{a=1}^{3\mathsf{g}-3}   \longrightarrow   \{G^{-1}_{t(a)}    g_{t(a) s(a)} G_{s(a)}  , G^{-1}_{t(a)} h_{t(a)} G_{t(a)}\}_{a=1}^{3\mathsf{g}-3} \; .
\end{equation}
As before, it is always possible to preserve the gauge fixing for the holonomy going from $n^{\cyl}_a$ to the  marked point of the target puncture.
To summarize, we now have a parametrization for a flat graph connection by equivalence classes
\ba\label{Param2}
(G^{6\mathsf{g}-6})_{|\text{flat}} / G^{2\mathsf{g}-2}
\ea
where the holonomy configurations have to satisfy the flatness constraints \eqref{Flatness2} and  the gauge orbits  are defined in \eqref{Gauge2}.

This description can be easily modified by contracting a cylinder, which connect two different spheres, and thus joining these two spheres  to obtain  a sphere with more punctures.  A contraction of a cylinder removes a $(g_a,h_{t(a)})$-pair from the configuration space---which is consistent as one has also reduced the number of spheres, and thus the number of flatness constraints and gauge actions, by one. We leave it to the reader to show that a contraction of cylinders along a maximal spanning tree (of the spine graph defined by taking the spheres as nodes and cylinders as edges) recovers consistently the minimal holonomy description in \eqref{Param1}.

\subsection{Towards the Drinfel'd Double parametrization}\label{tDDP}

The holonomy parametrization as given above is not free in the sense that many flatness constraints remain to be imposed on the holonomies and we still have to factor out the gauge action.  Starting from the holonomy parametrization which is based on a decomposition of the genus-$\mathsf{g}$ surface into thrice-punctured two-spheres $\mathbb{Y}$, we wish to fix the gauge action and at the same time solve for the flatness constraints associated with the spheres.  

Let us first consider the holonomies $\{h\}$ going around the punctures. The gauge invariance requires the states to be invariant under adjoint action on such holonomies. Therefore, we introduce a gauge invariant characterization in terms of the corresponding conjugacy classes. For each 
$\mathbb{Y}$, we have three clockwise-ordered punctures labeled by $( (a_1,o_1), (a_2,o_2), (a_3,o_3))$. The corresponding holonomies $(h_{o_1(a_1)},h_{o_2(a_2)} ,h_{o_3(a_3)})$ then satisfy the flatness constraint \eqref{Flatness2}. A gauge invariant characterization of this triple of holonomies is obtained by using their conjugacy classes $(C_{o_1(a_1)},C_{o_2(a_2)} ,C_{o_3(a_3)})$. The flatness constraint then select which triples of conjugacy classes are allowed to appear. In order to perform such a selection, it is convenient to introduce \emph{fusion coefficients} for the conjugacy classes. Let us consider the set
\be
	S =\{(h_{o_1(a_1)},h_{o_2(a_2)},h_{o_3(a_3)})\in C_{o_1(a_1)} \times C_{o_2(a_2)} \times C_{o_3(a_3)} \,| \, h_{o_3(a_3)}h_{o_2(a_2)}h_{o_1(a_1)} = \unitG \} \; 	.
\ee
The number of orbits the set $S$ splits into under simultaneous adjoint action defines fusion coefficients denoted by $N_{C_{o_1{a_1}}C_{o_2(a_2)}C_{o_3(a_3)}}$. These coefficients are non-vanishing only when the triple of conjugacy classes is admissible. As the name suggests, we can indeed understand this statement as a condition for the magnetic excitations labeled by $C_{o_1(a_1)}$ and $C_{o_2(a_2)}$ to fuse so as to obtain a magnetic excitation labeled by $C_{o_3(a_3)^{-1}}$ which is defined to be the conjugacy class of the inverse of any element of  $C_{o_3(a_3)}$. More precisely, the fusion coefficients count the number of inequivalent ways of satisfying such a fusion. In the following, we assume for notational convenience that for each allowed triple of conjugacy classes, there is only one representative triple of holonomies which satisfy the flatness constraint modulo a common adjoint action of the group. This amounts to assuming multiplicity freeness for the associated coupling. Replacing the $h$-holonomies by their conjugacy classes $C_h$, and allowing for each thrice-punctured sphere only triples of admissible conjugacy classes, is a first step towards a gauge invariant parametrization. Let us now analyze the remaining gauge freedom in more detail. 

Let us consider an (ordered) triple of conjugacy classes $(C_1,C_2,C_3)$ such that the corresponding fusion coefficients are non-vanishing and let us choose a representative (ordered) triple of holonomies $(h_1,h_2,h_3)\in C_1 \times C_2 \times C_3$ satisfying $h_3 h_2 h_1=\unitG$. We denote by $Z_{C_1C_2C_3}$ the  stabilizer group with respect to a simultaneous adjoint action of this representative triple. For each $\mathbb{Y}$, we equip the punctures with clockwise ordering and assign each ortiented cylinder with a conjugacy class so that the coupling conditions are satisfied. Assume that we have a configuration $ \{g_{t(a) s(a)}, h_{t(a)}\}_{a=1}^{3\mathsf{g}-3} $ consistent with this choice of conjugacy classes. We can use the gauge freedom at each $\mathbb{Y}$ to transform each consistent triple of holonomies $(h_{o_1(a_1)} , h_{o_2(a_2)} , h_{o_3(a_3)})$ into the representative triple of holonomies determined by the triple $(C_{o_1(a_1)} , C_{o_2(a_2)},  C_{o_3(a_3)})$ of conjugacy classes. This leaves us with a residual gauge freedom given by the stabilizer  groups $Z_{C_1C_2C_3}$ associated with every $\mathbb{Y}$.

So far we have been focusing on the $h$-holonomies, let us now consider the $g$-holonomies. More precisely, we want to determine how much freedom is left, after the above gauge fixing, in choosing the $g$-holonomies. The assignment of conjugacy classes to the cylinders, together with the above gauge fixing, determines uniquely all the $h_{s(a)}$- and $h_{t(a)}$-holonomies.  For a given cylinder $a$ with conjugacy class $C=C_a$, the choice of holonomy $g=g_a$  is restricted since the identity $h^{-1}_{s}:= g^{-1} h_{t}  g$  need to be satisfied so that it must take the form
\begin{equation}
	\label{gDouble}
	g\,=\,   q^{\phantom{-1}}_{\iota_C(h^{~}_t)} \cdot  z  \cdot q^{-1}_{\iota_C(h_s^{-1})}
\end{equation}
where $z \in Z_C$, the centralizer group of $C$, and $\iota_C$ is a labeling function such that $q_{\iota_C(h)}$ are elements of the quotient group $G/Z_C$ satisfying $c_1= q^{-1}_{\iota_C(h)} h q_{\iota_C(h)}$ with $c_1$ the representative of $C$.

Most importantly, it follows from the previous relation that for each cylinder $a$, the freedom left in choosing $g_a$ is parametrized by the stabilizer group $Z_{C_a}$ of the conjugacy class $C_a$ associated with the cylinder.  As noted above, we also have a residual gauge action given by the stabilizer groups $Z_{C_1C_2C_3}$ associated with each thrice-punctured two-sphere. Finding a completely gauge invariant parametrization for the remaining choice of the $g$-holonomies leads to the Drinfel'd double parametrization.

\subsection{The Drinfel'd double parametrization \label{sec:DDP}}

In sec.~\ref{sec:tube}, we derived a basis for excited states on the twice-punctured two-sphere labeled by Drinfel'd double irreducible representations. These states were obtained via a Drinfel'd double Fourier transform \eqref{cyl}. Furthermore, we explained earlier how in order to perform the gluing of surfaces, and of the states who live on such surfaces, it was necessary to apply a projector to impose flatness and gauge invariance. This gluing procedure is particularly simple for Drinfel'd double parametrized states \eqref{cyl} as it amounts to summing over the vector space indices of the representations at hand. The same strategy holds for more general states.
	
The Drinfel'd double (or fusion) basis states on the 	cylinder are labeled by $\rho=(C,R)$ where $C$ is the conjugacy class of the holonomy going around the cylinder and $R$---denoting an irreducible representation of the stabilizer group $Z_C$---characterizes the states dependence on the holonomy along the cylinder. In the previous section, we also found that the holonomies around the cylinders can be characterized by their conjugacy class $C$ and that the remaining freedom in choosing the $g$-holonomies along the cylinders is parametrized by the stabilizer groups $Z_C$ associated to each of the cylinders. This suggests that we should indeed assign to each cylinder a Drinfel'd double state and that we should glue or `fuse' these states at the thrice-punctured spheres connecting the cylinders. This fusion does indeed correspond to a fusion (or tensor) product in the category $\text{Rep}[\mathcal{D}(G)]$ 	of finite dimensional representations of the Drinfel'd double (see  app.~\ref{app_drinfeld} for more detailed definitions). That is, we can form the tensor product of two representations  
\begin{equation}
	\rho_1 \otimes \rho_2 = \bigoplus_{\rho_3} N^{\rho_3}_{\rho_1 \rho_2}\, \rho_3 \; .
\end{equation}
Furthermore, there exist Clebsch-Gordan coefficients $C^{\rho_1\rho_2\rho_3}_{N_1N_2N_3}$ satisfying
\begin{equation}
	\label{defCGC}
	D^{\rho_1}_{M_1N_1}\otimes D^{\rho_2}_{M_2N_2}(\Delta(g \smo \delta_h)) = \sum_{\rho_3}\sum_{M_3N_3}
	\mathcal{C}^{\rho_1\rho_2\rho_3}_{M_1M_2M_3}\, D^{\rho_3}_{M_3N_3}(g \smo \delta_h)\,\overline{\mathcal{C}^{\rho_1\rho_2\rho_3}_{N_1N_2N_3}} 
\end{equation}
where $\Delta$ denotes the comultiplication map (see app.~\ref{app_drinfeld}). Such coefficients act as an intertwiner between $V_{\rho_1} \otimes V_{\rho_2}$ and $V_{\rho_3}$. It is often convenient to use the more symmetric definition
\begin{equation}
	\Big( {}^{\, \rho_1 \; \rho_2 \; \rho_3}_{N_1N_2N_3}\Big)
	:= \frac{1}{\sqrt{d_{\rho_3}}}C^{\rho_1\rho_2\rho_3^\ast}_{N_1N_2N_3} \; ,
\end{equation}
where $\rho_3^\ast$ denotes the representation dual to $\rho_3$, that define the mapping $V_{\rho_1}\otimes V_{\rho_2} \otimes V_{\rho_3} \rightarrow V_0$. By analogy with the group case, we refer to these coefficients as the $3 \rho M$-symbols.

By definition the $3 \rho M$-symbols satisfy an invariance property. Let us for instance consider the situation where the $3 \rho M$-symbols perform the mapping $V_{\rho_1}^\ast \otimes V_{\rho_2} \otimes V_{\rho_3}^\ast \rightarrow V_0$. In this case, the invariance of the $3 \rho M$-symbols reads
\begin{equation}
\label{invCG}
\frac{1}{|G|}\sum_{h_1,h_2}
D^{\rho_1^\ast}_{M_1N_1}(g \smo \delta_{h_1})D^{\rho_2}_{M_2N_2}(g \smo \delta_{h_2})
D^{\rho_3^\ast}_{M_3N_3}(g \smo \delta_{h_2^{-1}h_1^{-1}})\Big({}^{\, \rho_1^\ast \; \rho_2 \; \rho_3^\ast}_{N_1N_2N_3}\Big)
= \Big({}^{\, \rho_1^\ast \; \rho_2 \; \rho_3^\ast}_{M_1M_2M_3}\Big)\; .
\end{equation}
This can be used to show that the $3 \rho M$-symbols implement the flatness constraint and the gauge invariance when gluing three cylinders states to a thrice-punctured sphere (see \cite{DDR1} for details). For example, fig.~\ref{fig_3puncsphere} displays the gluing associated to the configuration of \eqref{invCG} where the states $| \rho_1^\ast \ra_{\cyl}$ and $| \rho_3^\ast \ra_{\cyl}$ are ingoing while the state $| \rho_2 \ra_{\cyl}$ is outgoing. By convention, the $3 \rho M$-symbols $\Big({}^{\, \rho_1 \; \rho_2 \; \rho_3}_{N_1N_2N_3}\Big)$ are associated to three outgoing states labeled by $\rho_1$, $\rho_2$ and $\rho_3$. This explains why we consider the dual representations $\rho_1^\ast$ and $\rho_3^\ast$ in our example. This gluing of three cylinder states via the $3 \rho M$-symbols provides us with the following fusion basis states for the thrice-punctured two-sphere $\mathbb{Y}$:
\be
	|\rho_1^\ast,N_1 ; \rho_2, M_2 ; \rho_3^\ast,N_3 \ra_{\mathbb{Y}} :=
	\sum_{M_1,N_2,M_3}
	| \rho_1^\ast,M_1N_1 \ra_{\cyl} \otimes | \rho_2,M_2N_2 \ra_{\cyl} \otimes
	| \rho_3^\ast, M_3N_3 \ra_{\cyl} \otimes 	\Big({}^{\, \rho_1^\ast \; \rho_2 \; \rho_3^\ast}_{M_1N_2M_3}\Big)  \; .
\ee 
From the discussion above, it follows that the flatness (or Bianchi) constraint for the sphere is satisfied and that the gauge invariance at the internal nodes is implemented. Therefore, in order to obtain the fusion basis states  for a genus-$\mathsf{g}$ surface $\Sigma$ we proceed as follows:
$(i)$ Choose a pant decomposition $\{\mathbb{Y}\}$ of the surface $\Sigma$. $(ii)$ Associate with each $\mathbb{Y}$ a fusion basis state. $(iii)$ Glue all the fusion basis states together by summing over the corresponding vector space indices according to the pattern of the pant decomposition. In this way, we associate also to each cylinder  in the pant decomposition a Drinfel'd double irreducible representation $\rho=(C,R)$. We have furthermore coupling conditions for each triple of representations $\rho_1,\rho_2,\rho_3$  meeting at a sphere which do entail the coupling conditions for the conjugacy classes, that we discussed in the previous section.

The $C$-labels therefore agree with the  parameterization discussed in the previous section. The remaining $R$-labels denote irreducible representations of the stabilizer groups $Z_{C}$ of $C$. Indeed, we saw in the previous section that the remaining freedom in the $g$-connections is parametrized by the stabilizer groups $Z_{C}$, but that there is also a remaining gauge freedom given by the stabilizer groups $Z_{C_1C_2C_3}$ associated to each thrice-punctured sphere. But the  Drinfel'd double basis does encode the $g$-connection into a  spin network, with representation labels conditioned by the $C$-labels. This spin network description  ensures gauge invariance with respect to the remaining gauge freedom given by the stabilizers $Z_{C_1C_2C_3}$. Note finally that this construction also applies to 2d  surfaces with boundary since it is still possible to obtain them as a gluing of thrice-punctured two-spheres $\mathbb{Y}$.

\newpage
\section{Lifting procedure to the (3+1)d case \label{sec:lifting}}
We can now construct the state spaces for flat connections on manifolds ${\cal M}$ with a defect structure. The first step of this construction consists in applying the previous procedure to define the Hilbert space of flat connections $\mathcal{H}_{\Sigma_{\mathsf{H}}}$ on a two-dimensional surface $\Sigma_{\mathsf{H}}$, obtained from a Heegaard splitting of a three-manifold $\mathcal{M}$, adjusted to the defect structure.  By imposing further constraints on the states in $\mathcal{H}_{\Sigma_{\mathsf{H}}}$, we obtain a Hilbert space of flat connections which can describe magnetic (or curvature) excitations attached to the chosen defect structure. Here we assume that the defect structure is given by the one-skeleton of a triangulation. The construction is however easily generalizable to other lattices.

\subsection{Heegaard splitting} 

A \emph{Heegaard splitting} \cite{gompf19994} is a decomposition of a compact three-manifold $\mathcal{M} $  into two handlebodies $\mathcal{M}_1$ and $\mathcal{M}_2$ such that $\mathcal{M}= \mathcal{M}_1 \cup \mathcal{M}_2$. This splitting is performed along the so-called Heegaard surface $\Sigma_\mathsf{H}$, that is $\Sigma_\mathsf{H}$ is the boundary of each of the two handlebodies $\partial \mathcal{M}_1 = \partial \mathcal{M}_2 = \Sigma_\mathsf{H}$. 

One way of obtaining such a splitting is via a triangulation. Let $\large\triangle$ be a trianguation of $\mathcal{M}$ and $\large\triangle_1$ its one-skeleton, i.e. the union of its vertices and edges. We then consider the regular neighbourhood $\pmb{\large\triangle}_1$ of $\large\triangle_1$ obtained by blowing-up the vertices and edges into a union of 3-balls and solid cylinders, respectively. This regular neighbourhood provides the first handlebody $\mathcal{M}_1 \simeq \blow$ and $\mathcal{M}_2$ is defined as its complement in $\mathcal{M}$  i.e. as $\mathcal{M} \backslash \blow$. The surface of this regular neighbourhood defines the Heegaard surface $\Sigma_\mathsf{H}(\large\triangle)$ associated with the triangulation $\large\triangle$. Similarly, we can consider the regular neighbourhood of the dual graph $\Upsilon$ to the triangulation, which is homeomorphic to $\mathcal{M}_2$ defined above, that is the complement of the regular neighbourhood of the one-skeleton $\blow$ in $\mathcal{M}$. The surface of the regular neighbourhood of the dual graph is homeomorphic to $\Sigma_\mathsf{H}(\large\triangle)$.

Let us now consider a graph $\Gamma$ embedded on $\Sigma_\mathsf{H}(\large\triangle)$ which captures the non-contractible cycles of the fundamental group $\pi_1(\Sigma_\mathsf{H}(\triangle))$ of $\Sigma_\mathsf{H}(\large\triangle)$.  
We distinguish on the Heegaard surface $\Sigma_\mathsf{H}(\large\triangle)$ two sets of closed curves, which we denote by $\{\mathcal{C}_t\}$ and $\{\mathcal{C}_e\}$. 
The first set  $\{\mathcal{C}_{t}\}_{t \subset \large\triangle}$ consists of the curves around the triangles, that is each triangle $t$ contributes a curve $t \cap \Sigma_\mathsf{H}(\large\triangle)$. The second set $\{\mathcal{C}_e\}_{e \subset \large\triangle} $ is given by the curves around the edges $e$ of the triangulation, that is for each edge we choose a disk $d_e$ (of appropriate size) intersecting the edge $e$ transversally and consider the curve $d_e \cap \Sigma_\mathsf{H}(\large\triangle)$. The set $\{\mathcal{C}_t\}_{t \subset \triangle}$ generates all curves that are contractible in $\mathcal{M} \backslash \blow$ but are not contractible in $\blow$. Note that, as far as this generating property is concerned, the set $\{\mathcal{C}_{t}\}_{t \subset \triangle}$ is in general over-complete. In terms of the holonomies associated with these curves, this over-completeness leads to one Bianchi identity for each of the 3-simplices of $\triangle$. (The Bianchi identities may themselves be over-complete.) Conversely, the set  $\{\mathcal{C}_{e}\}_{e \subset \large\triangle}$ generates all curves that are contractible in $\blow$, but not in $\mathcal{M} \backslash \blow$. Again, this set is often over-complete which now leads to Bianchi identities associated with the vertices of the triangulation. 

Let us now describe the space of flat connections on ${\cal M}\backslash \blow$ using the Heegaard surface  $\hee$: We do so by considering the Hilbert space ${\cal H}_{\hee}$ and impose on this space additional flatness constraints. These flatness constraints demand that the holonomies along the curves in $\{\mathcal{C}_{t}\}$ are trivial. In the following, we refer to the set of flatness constraints associated with the curves in $\{\mathcal{C}_{t}\}$ as \emph{two-handle constraints}.\footnote{This is reminiscent of the fact that when blowing-up the one-skeleton of the triangulation $\triangle$, the blown-up edges form one-handles while the blown-up triangles form two-handles.} Thus given a Hilbert space ${\cal H}_{\hee}$ of wave functions on the space of flat connections on $\hee$, we can obtain a Hilbert space of wave functions of flat connections on ${\cal M}\backslash \blow$ by defining suitable operator versions of the two-handle constraints and projecting onto the subspace of functions satisfying these constraints. This subspace defines ${\cal H}_{\mathcal{M} \backslash \blow}$.\footnote{ 
 This task is straightforward in the finite group case as the subspace of states satisfying the constraints is a proper subspace of ${\cal H}_{\hee}$. For Lie groups the two-handle constraints are given by delta functions on the group. If one chooses a measure for ${\cal H}_{\hee}$ constructed from the Haar measure on the group,  wave functions satisfying these constraints will not be normalizable with respect to the inner product of ${\cal H}_{\hee}$. Thus the space of solutions has to be equipped with a new inner product. Several methods are available \cite{RAQ1,MCP1,MCP2} for the construction of such inner products. We will see that we regain the spin network basis, which is also well defined for the Lie group case and thus we expect that a suitable procedure can be found. Another possibility is to consider a discrete measure for ${\cal H}_{\hee}$, which in fact arises if one wishes to work with continuum Hilbert spaces \cite{DGflux, DGfluxQ,Lewandowski:2015xqa}. 
In this case the spectrum of the two-handle constraints will be discrete and thus the solutions to the constraints will be normalizable.}

\begin{figure}
	\centering
	\label{fig_tetra}
	\includegraphics[scale=0.85,valign=c]{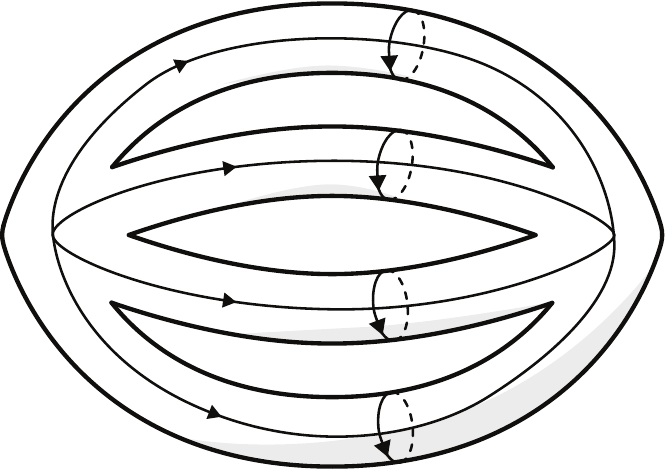} \q 
	\includegraphics[scale=0.85,valign=c]{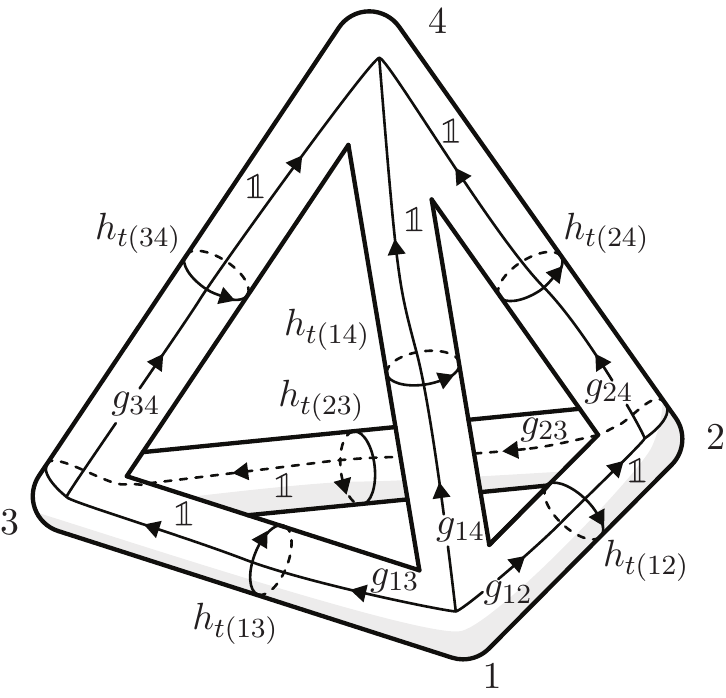}
	\caption{On the right panel, we consider the simplest triangulation of the three-sphere manifold $\mathcal{M}$, namely two tetrahedra identified to each other. By blowing-up the one skeleton of this triangulation, we obtain a Heegaard splitting $\mathcal{M} = \blow \cup \mathcal{M} \backslash \blow$. On the Heegaard surface $\Sigma_\mathsf{H}(\large\triangle)$, we construct a graph which capture all the non-contractible cycles on the surface. The cycles $\{\mathcal{C}_{e}\}_{e \subset \triangle}$, which go around the edges of the triangulation, are contractible in $\blow$ while the cycles $\{\mathcal{C}_t\}_{t \subset \triangle}$, which run along the edges of $\large\triangle$ and are homotopic to the triangles, are contractible in $\mathcal{M} \backslash \blow$. This means in particular that the line defects for $\mathcal{M} \backslash \blow$ run along $\{\mathcal{C}_{e}\}_{e \subset \triangle}$. On the left panel, we represent the blown-up of the one skeleton $\Upsilon_1$ of the dual cell-complex and define a graph on its surface in a similar fashion. The four-times-punctured sphere can be further split into two thrice-punctured sphere so as to obtain a pant decomposition.}
\end{figure}

We discussed in sec.~\ref{tDDP} that for the definition of a fusion basis on $\hee$, we need to choose a pant decomposition for this surface.  In the following, we consider two classes of such pant decompositions, which are adjusted to the one-skeleton $\triangle_1$ of the triangulation and the one-skeleton $\Upsilon_1$  of the dual cell-complex, respectively.

\subsection{The spin network basis}

 In this section, we consider the case of a pant decomposition adjusted to $\Upsilon_1$, that is the one-skeleton of the dual complex to the triangulation. We show that in this case the two-handle constraints just pick out a subset of the fusion basis states in ${\cal H}_{\hee}$, and that this subset agrees with the (gauge invariant) spin network basis \cite{Rovelli:1995ac} for lattice gauge theory and loop quantum gravity.

Such a pant decomposition can be obtained by first cutting the surface $\Sigma_\mathsf{H}(\large\triangle)$ along the triangle curves $\{\mathcal{C}_{t}\}_{t \subset \large\triangle}$.  This decomposes the surface into four-punctured spheres, each one of them being associated with a tetrahedron of the triangulation. These spheres are connected by cylinders that surround the edges of the dual graph $\Upsilon_1$. Each four-punctured sphere can then be decomposed---in one of the three possible ways---into two thrice-punctured spheres in order to obtain a pant decomposition. The fusion basis associated with such a pant decomposition assign labels $(C_{e^*},R_{e^*})$  to each dual edge $e^*=t$. Furthermore, to each vertex $v^\ast$ dual to a 3-simplex of the triangulation, we assign a pair of labels $(C_{v^*},R_{v*})$ which are associated to the \emph{virtual} cylinders connecting the pairs of thrice-punctured spheres resulting from decomposing the four-times-punctured ones.

The $C_{e^*}$-labels provide the conjugacy classes associated with the holonomies going around the dual edges, that is around the triangles of $\triangle$. This means that the two-handle constraints are satisfied if and only if $\{C_{e*}=\{\unitG\}\}$ for all the dual edges $e^*$. Imposing these conditions, each of the thrice-punctured spheres will have two punctures carrying labels with a trivial conjugacy class. It then follows from the fusion coefficients for the conjugacy classes that the conjugacy classes associated with the remaining punctures must be trivial as well, i.e. $\{C_{v^*} =\{\unitG\}\}$ for all dual vertices $v^\ast$.\footnote{The argument generalizes also to lattice with  three-dimensional building blocks which have more than four faces: A three-dimensional building block with $n$ faces is dual to an $n$-valent vertex. This $n$-valent vertex can be extended into a three-valent  tree graph with $n$ leaves, which represent the faces of the building block  and $(n-3)$ internal edges which represent virtual cylinders. The two-handle constraints impose a trivial conjugacy class for all leaves. The fusion rules then impose that the conjugacy classes for the virtual cylinders are also trivial.}

Since all the conjugacy classes are trivial, the corresponding stabilizer groups always coincide with the full group $G$ and {\it a fortiori} the representation labels $\{R\}$ stand for irreducible representations of this group. In summary, we have $G$-representation labels $\{R_{e^*}\}$ associated with the edges of the dual graph together with representation labels $\{R_{v^*}\}$ associated with the dual vertices. This latter set can be interpreted as labels for the four-valent intertwiners associated with the dual vertices, that is the tetrahedra of the triangulation.\footnote{Remember that we assumed multiplicity freeness for the tensor product of two Drinfel'd double representations, which implies multiplicity freeness for the tensor product of two group representations.} We have therefore reconstructed the spin network basis \cite{Rovelli:1995ac}.  This can be easily confirmed by considering the Drinfel'd double Fourier transform defined in \eqref{cyl} for trivial conjugacy classes. Indeed, it follows from equations \eqref{defReps} and \eqref{defCGC} that the irreducible representations and the Clebsch-Grodan coefficients for $\mD(G)$ reduce to the ones for the group $G$ if all the conjugacy classes are trivial.

\subsection{The dual spin network basis \label{sec:curv}} 
Let us now consider the case of a pant decomposition adjusted to $\triangle_1$, that is the one-skeleton of the triangulation. In this case, the projection procedure is far more involved. A pant decomposition associated with $\triangle_1$ can be obtained by cutting the Heegaard surface along the set of curves $\{\mathcal{C}_e\}_{e\subset \large\triangle}$.  By doing so, we associate with each $m$-valent vertex of the triangulation an $m$-times-punctured sphere. We can again freely decompose these $m$-times-punctured spheres into thrice-punctured ones. In order to discuss the imposition of the two-handle constraints, we make use of the parametrization of the space of flat connections developed in sec.~\ref{holp} and \ref{tDDP}.  

Let us first briefly recall the definition of the holonomies used in the parametrization of the space of flat connections. In particular, we want to clarify their meaning for the pant decomposition at hand. We start from the picture where to each oriented cylinder $a$ appearing in the pant decomposition we associated a holonomy $h_{t(a)}$ and a holonomy $g_a=g_{t(a)s(a)}$. This determines another holonomy associated with the source puncture of the cylinder whose expression reads $h_{s(a)}=g_a^{-1} h^{-1}_{t(a)}g_a$. A subset of cylinders can be identified with the edges of the triangulation. We choose an orientation for these edges and denote the corresponding cylinders by $e$ instead of $a$. The remaining cylinders are associated with the vertices of the triangulation. For each $m$-valent vertex of $\triangle$, we  have $(m-3)$ pairs of variables $(h_{t(v,j)},g_{v,j})_{j=1}^{m-3}$  arising from the decomposition of the associated $m$-punctured spheres into thrice-punctured ones. 
The $h_{t(v,j)}$ variables give the holonomies around certain sets of edges starting from the same vertex $v$. The $g_{v,j}$ variables complete the information  about the parallel transport on the punctured spheres, so that the set of $g$-holonomies provide a graph connection along the blown-up one-skeleton $\blow$. Note that is important to keep track of how the $g$-holonomies wind around the punctures of a given sphere. 

The set of holonomies we just described is supposed to define a flat connection on $\hee$. This set is therefore subject to constraints that are given in terms of the Bianchi identities for the thrice-punctured spheres.   As explained in sec.~\ref{tDDP}, the Bianchi identities are imposed  by associating to each cylinder $a=e$ or $a=(v,j)$  conjugacy classes $C_a$ so that the coupling conditions for the conjugacy classes are satisfied. Furthermore, we need to enforce the two-handle constraints in order to project the states in ${\cal H}_{\hee}$ onto the Hilbert space ${\cal H}_{\mathcal{M} \backslash \blow}$.  In order to identify which holonomies are set to the identity, we need to consider, for each triangle $t \subset \triangle$, the triangle curve ${\cal C}_t$ and isotopically deform this curve so that it matches a path along the chosen graph $\Gamma$ on $\hee$.   Since the triangle curves necessarily go along the edges of the triangulation, the two-handle constraints must involve the $g$-holonomies.  However, in order to obtain paths which are isotopically equivalent to the triangle curves, it might be necessary to include some windings around the punctures. In such cases, the two-handle constraints also involve some $h$-holonomies. 

We described in sec.~\ref{tDDP} how, given a consistent set of conjugacy classes $\{C_a\}_a$, we can construct a partial gauge fixing which determines uniquely all the $h$-holonomies. This partial gauge fixing is such that $g_a$-holonomies are restricted to be of the form
\begin{equation}
	\label{Formg2}
	g_a\,=\,    q^{\phantom{-1}}_{\iota_{C_a}(h_t)} \cdot   z_a \cdot  q^{-1}_{\iota_{C_a}(h_s^{-1})} \equiv q_{\iota_a}^{\phantom{-1}}\!\!\!  \cdot   z_a  \cdot q^{-1}_{\iota'_a}
\end{equation}
where the remaining freedom is parametrized by $z_a\in Z_{C_a}$ and $q_{\iota_C(h)}$ is defined as in \eqref{gDouble}. Henceforth, we will make use of the shorthand notation $q_{\iota_a} \equiv q_{\iota_{C_a}(h_t)}$ and $q_{\iota'_a} \equiv q_{\iota_{C_a}(h_s^{-1})}$. Recall finally that when performing the partial gauge fixing, for a given consistent set of conjugacy classes $\{C_a\}_a$, there is some remaining gauge freedom given by the stabilizer groups $Z_{C_1C_2C_3}$ associated with the thrice-punctured spheres. For a given configuration $\{C_a\}_a$ of conjugacy classes, we thus understand the two-handle constraints as (flatness) conditions on the variables $\{z_a \in Z_{C_a}\}_a$. We distinguish three possible scenarios:
\begin{enumerate}[itemsep=0.4em,parsep=1pt,leftmargin=*]
	\item[${\sss (1)}$] The two-handle constraints admit no solution, in which case we have to exclude the configuration $\{C_a\}_a$. 
	In this case we have to conclude that a basis state with a curvature configuration $\{C_a\}_a$ does not exist.
	
	\item[${\sss (2)}$] The two-handle constraints admit a unique solution modulo the gauge action given by the stabilizer groups $Z_{C_1C_2C_3}$ associated to the thrice-punctured spheres. In this case we can conclude that there is a unique state peaked on a curvature configuration $\{C_a\}_a$.
	
	\item[${\sss (3)}$] There are several (left-over gauge) orbits satisfying the two-handle constraints so that we need to introduce additional quantum numbers $\{Q_I\}_I$ which label such orbits. 
For instance, one can define closed holonomy operators ${\cal O}_I$, i.e. operators measuring the conjugacy class of holonomies associated to certain loops, that differentiate between these orbits.
\end{enumerate}

\medskip \noindent
One would expect the scenario ${\sss (3)}$ to occur for manifolds ${\cal M}$ with non-trivial $\pi_1({\cal M})$, in particular for the case of vanishing magnetic excitations $\{C_a=\{\unitG\}\}_a$ for all $a$.  In this case, we should still be left with the space of flat connections described by homomorphisms of $\pi_1({\cal M})$ into $G$ (modulo the adjoint action). Conversely, for a trivial $\pi_1({\cal M})$, one would find a unique basis state with configuration $\{ C_a=\{\unitG\}\}$. We will however see that the cases ${\sss (3)}$, and  ${\sss (1)}$ do appear also for triangulations of the three-sphere. In the case where ${\sss (1)}$ appears, we can interpret the absence of the corresponding configuration of conjugacy classes $\{C_a\}$ as an additional coupling rule. 

In order to distinguish the different solutions appearing in case ${\sss (3)}$, we have to add additional (gauge invariant) observables $\{{\cal O}_I\}_I$. These can be chosen to be conjugacy classes of cycles which involve the $g$-holonomies. For instance, if $\mathcal{M}$ has non-trivial topology, we can choose cycles along the Heegaard surface that generate $\pi_1(\mathcal{M})$, and we can pick as observables the conjugacy classes of the holonomies associated to these cycles as well as products of these cycles.  One can however also construct more local observables that capture gauge invariant information about the $g$-holonomies.  Let us assume that there are solutions to the two-handle constraints that differ in their $g_a$ value for a particular cylinder $a$, which goes between two spheres $s(a)$ and $t(a)$.  In this case, we can consider the conjugacy class of the holonomy $g_a^{-1} h_T g_a  h_S$, where $h_S$ is a holonomy around a cylinder $b$ adjacent to the sphere $s(a)$ and $h_T$ is a holonomy around a cylinder $c$ adjacent to the sphere $t(a)$. The basis states can then be expressed as
\ba
| \{C_a\}_a, \{C_I\}_I \rangle \,=\, \frac{1}{{\cal N}(\{C\}) }   \sum_{\{g_a,h_a\}} \prod_a  \Theta_{C_a}(h_a)  \,\,  \prod_I \Theta_{C_I}\big({\cal O}_I(\{g_a,h_a\} )\big) \,\, |\{g_a,h_a\}_a\rangle 
\ea
where $\Theta_C( {\sss \bullet})$ is the characteristic function of $C$ and ${\cal N}(\{C\})$  a suitable normalization constant.

\subsection{Examples}

In this section, we present several examples of our construction based on Heegaard splittings. In order to perform the computations explicitly, we consider the non-abelian group of permutations of three elements denoted by ${\cal S}_3$. 

\subsubsection{Preliminary: Symmetric group $\mathcal{S}_3$}

The symmetric group $\mathcal{S}_3$ is the simplest example of finite non-abelian group. It is the symmetry group of an equilateral triangle. This group is generated by a rotation with a $2\pi/3$ angle as well as a reflection with respect to any of the three medians. The groups associated with these two generators are the cyclic groups $\mathbb{Z}_3$ and $\mathbb{Z}_2$, respectively. We denote the generators of $\mathbb{Z}_2$ and $\mathbb{Z}_3$ by $r$ and $s$, respectively, such that $r^2 = \unit$ and $s^3 = \unit$. The six group elements of $\mathcal{S}_3$ are therefore given by
\begin{equation}
	{\cal S}_3 = \{r^is^j\}_{i=0,1}^{j=0,1,2} =  \{ \unit, r, rs, rs^2, s ,s ^2\} \; .
\end{equation}
Using the defining relation of the generators $r$ and $s$ together with the relation $rs = s^2r$, we obtain the following multiplication table:

\begin{equation*}
\begin{tabular}{ l|c c c c c c } 
	$\cdot$ & $\unit$ & $r$ & $rs$ & $rs^2$ & $s$ & $s^2$ \\
	\hline
	$\unit^{\phantom{\big|}}$ & $\unit$ & $r$ & $rs$ & $rs^2$ & $s$ & $s^2$ \\
	$r$ & $r$ & $\unit$ & $s$ & $s^2$ & $rs$ & $rs^2$ \\
	$rs$ & $rs$ & $s^2$ & $\unit$ & $s$ & $rs^2$ & $r$ \\
	$rs^2$ & $rs^2$ & $s$ & $s^2$ & $\unit$ & $r$ & $rs$ \\
	$s$ & $s$ & $rs^2$ & $r$ & $rs$ & $s^2$ & $\unit$ \\
	$s^2$ & $s^2$ & $rs$ & $rs^2$ & $r$ & $\unit$ & $s$ 
\end{tabular}
\end{equation*}
\noindent
The group elements can be classified according to whether they are \emph{odd} or \emph{even}, that is, whether their expression contains an odd number of $r$ elements or an even number. Correspondingly the determinant in the fundamental representation is equal to $-1$ for odd elements and equal to $+1$ for even elements. There are only three distinct conjugacy classes 
\begin{equation}
	{\cal T}=\{\unit\} \q , \q  {\cal O}=\{ r, rs,rs^2 \} \q , \q {\cal E}=\{s,s^2\}  \; ,
\end{equation}
the trivial conjugacy class ${\cal T}$, the conjugacy class ${\cal O}$ containing all odd elements and the conjugacy class ${\cal E}$ containing the even elements except for the unit $\unit$. The corresponding stabilizer groups are
\begin{align}
	Z_{\cal T}& \simeq{\cal S}_3 \; ,\nn\\
	Z_{\cal O}&= \{\unit,r\}\,\simeq\, \mathbb{Z}_2 \; , \nn\\
	Z_{\cal E}&=\{ \unit,s,s^2\}\,\simeq\, \mathbb{Z}_3
\end{align}
which are defined with respect to the following conjugacy class representatives: $c_1({\cal T})=\unit$, $c_1({\cal O})=r$ and $c_1({\cal E})=s$.

Recall that the fusion coefficients for the conjugacy classes are defined as the number of orbits the set
\begin{equation*}
	\{(h_1,h_2,h_3)\in C_1 \times C_2 \times C_3 \, | \, h_3h_2h_1 = \unit\}
\end{equation*}
splits into under simultaneous adjoint action. For ${\cal S}_3$ there are 11 non-vanishing fusion rule coefficients for which there exists a triple $(h_1,h_2,h_3)$ of group elements with $h_i\in C_i$ and $h_3h_2h_1=\unit$.  Furthermore, for the group ${\cal S}_3$  it holds that for a given $(C_1,C_2,C_3)$ all such triples of group elements are related by a common adjoint action transformation. Therefore, there is a one-to-one correspondence between each ordered triple $(h_1,h_2,h_3)$ satisfying the flatness condition and the corresponding triple of allowed conjugacy classes $(C_1,C_2,C_3)$. We summarize in the following table these 11 configurations:

\begin{equation}
	\label{couplingC}
	\begin{tabular}{ l|c c c } 
	$\otimes$ & $\mathcal{T}$ & $\mathcal{O}$ & $\mathcal{E}$ \\
	\hline
	$\mathcal{T}^{\phantom{\big|}}$ & $\mathcal{T}$ & $\mathcal{O}$ & $\mathcal{E}$ \\
	$\mathcal{O}$ & $\mathcal{O}$ & $\mathcal{T} \oplus \mathcal{E}$ & $\mathcal{O}$  \\
	$\mathcal{E}$ & $\mathcal{E}$ & $\mathcal{O}$ & $\mathcal{T} \oplus \mathcal{E}$ \\
	\end{tabular} 
\end{equation}
\noindent
Let us finally discuss the irreducible representations of ${\cal S}_3$. There are three irreducible representations. This includes the trivial one $R=0$. The sign representation $R=1$ is one-dimensional and gives $D^{R=1}(g)=\pm 1$ depending on whether $g$ is even or odd. Then there is the two-dimensional defining representation $R=2$, which derives from the group of rotations and reflections $O(2)$ in two dimensions.  There are also 11 possible couplings between these representations as can be seen from the following table:

\begin{equation}
	\begin{tabular}{ l|c c c } 
	$\otimes$ & 0 & 1 & 2 \\
	\hline
	$0^{\phantom{\big|}}$ & 0 & 1& 2 \\
	1 & 1 & 0 & 2  \\
	2 & 2 & 2 & 0 $\oplus$ 1$\oplus$ 2 \\
	\end{tabular}   
\end{equation}

\subsubsection{Genus-2 defect}

\begin{figure}
	\centering
	\includegraphics[scale=0.85,valign=c]{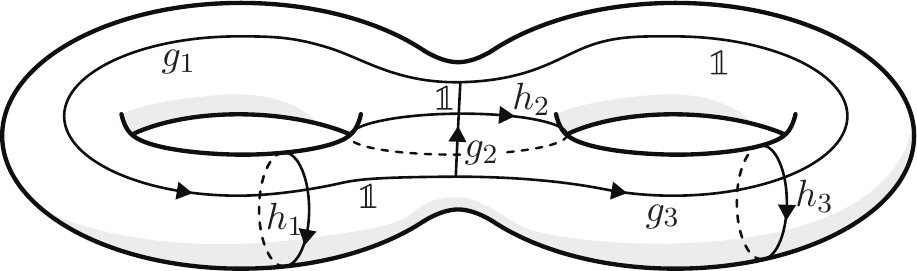}
	\caption{Example of embedded graph on a genus-2 surface.}
	\label{fig:doubleT}
\end{figure}

In sec.~\ref{sec:simple_ex}, we considered the simplest example of our construction, namely a loop defect embedded in the three-sphere. A slightly more complicated example consists in a genus-2 defect embedded in the three-sphere. 

Let us consider a graph embedded in $\mathbb{S}_3$ which consists of three edges and two vertices so as to form a $\theta$-shape. By blowing-up this graph we obtain a double torus which defines the Heegaard surface $\hee$.  The Heegaard surface splits the three-sphere into the blow-up $\blow$ of this graph, which gives a solid double torus, and the complement $\mathbb{S}_3 \backslash \blow$, which happens to be also a solid double torus. This defines the  genus-2 Heegaard splitting of the three-sphere. We represent in fig.~\ref{fig:doubleT} the Heegaard surface together with an embedded graph.

We define a graph connection for the graph in fig. \ref{fig:doubleT} by specifying three pairs of group elements $\{(g_a,h_a=h_{t(a)})\}_{a=1}^3$. The group elements for the remaining links are gauge fixed to $\unit$. To obtain a flat connection, we need to impose the Bianchi identities for the two spheres. We can use one of the Bianchi identities to solve for $h_3$ so that we are left with the following \emph{global} Bianchi identity
\ba\label{BianchiDT}
   h_2^{-1} g_2^{-1} h_2 g_2 \,g_3^{-1}h_3 g_3  h_3^{-1} \,=\, \unit  \; .
\ea
We can furthermore gauge fix $g_3=\unit$ which leaves us with a residual gauge action, given by the diagonal adjoint action on the configurations $(g_1,h_1,g_2,h_2)$. The two-handle constraints
$
	g_1g_2\,=\, \unit$  and  $g_2^{-1}g_3 \,=\, \unit
$
fix $g_1=\unit$ and $g_2=\unit$.  These solutions do also satisfy the Bianchi identity \eqref{BianchiDT}. Consequently, we are only left with the parameters $(h_1,h_2)$ which are subject to a diagonal adjoint action.  The corresponding 11 (for $\mathcal{S}_3$) equivalence classes are in one-to-one correspondence with the admissible triples of conjugacy classes $(C^*_1,C_2,C_3)$ (considering the vertex where the $a=1$ cylinder is outgoing). These conjugacy classes encode the values for the Wilson loop observables going around each of the edges of the defect structure given by the $\theta$-graph.

We noted above that the complement of the double torus in the three-sphere is again a double torus. Therefore, the corresponding dual graph is also given by a $\theta$-graph.  We can use a similar parametrization of the space of flat connection on the surface surrounding this dual graph as before, but now the two-handle constraints demand that $h_1=h_2=h_3=\unit$.  We are thus left with the $g$-holonomies, for which we have to impose gauge invariance. Using the corresponding Drinfel'd basis, the two-handle constraints impose that all conjugacy classes are trivial. We remain with the $R$-labels, which are now irreducible representations of the full group, and define a spin network for the $\theta$-graph. This spin network basis does indeed give a complete gauge invariant basis for the space of $g$-connections.  For the group $\mathcal{S}_3$, there are 11 spin network states for the $\theta$-graph, which is consistent with the number of states in the dual spin network basis.

\subsubsection{Defects along a tetrahedral skeleton \label{sec:ex_tetra}}

Let us now consider a tetrahedron $\triangle$ embedded into the three-sphere. We then perform a Heegaard splitting of the three-sphere such that the Heeagaard surface $\hee$ is obtained as the blow-up of the one-skeleton of the tetrahedron. The result is a genus-3 two-dimensional surface which we equip with a flat connection. By performing a pant decomposition, we obtain $\hee$ as a gluing of four $\mathbb{Y}$ associated with the vertices of $\triangle$ labeled by $k=1,\ldots, 4$. For each sphere, we choose a base node $n_k$. The spheres are glued to each other via tubes labeled by $a=21,31,41,32,42,43$, where we choose the sphere with the smaller label to be the source of the tube.  Each tube intersects its source and target $\mathbb{Y}$ at punctures which possess a marked point on their boundary. These marked points define the nodes $n_{s(a)}$ and $n_{t(a)}$. Between these nodes, we have a link going from $n_{s(a)}$ to $n_{t(a)}$. Furthermore, for each sphere $k$ we choose a link from the base node $k$ to all the $n_{s(a)}$ and $n_{t(a)}$ on this sphere. Finally, we choose links going around (clockwise on the punctured spheres) both ends of each tube that start and end at $n_{s(a)}$ and $n_{t(a)}$, respectively. Putting everything together, we obtain a graph $\Gamma$ on $\hee$. These conventions are summarized in fig.~\ref{fig_3puncsphere}.  

We can now define a flat connection on $\hee$ by assigning group elements $\{g_a\}$ to each link running along the cylinders $a$ from $n_{s(a)}$ to $n_{t(a)}$, as well as holonomies $\{h_{t(a)}\}$ going along the links encircling the target punctures of the cylinders. The holonomies from $n_k$ to the adjacent $n_{s(a)}$ and $n_{t(b)}$ are gauge fixed to the identity by using the gauge freedom at these latter nodes. 

Following this procedure, we obtain a parametrization which depends on six pair of group elements $\{g_a,h_{t(a)}\}$. This parametrization is over-complete since, on the one hand, there is a remaining gauge freedom for the base nodes $\{n_k\}$ and, on the other hand, the Bianchi identity holds at each $\mathbb{Y}$. To express this systematically, it is convenient to introduce the holonomies around the source puctures of the tubes, namely $\{h_{s(a)}= g_{a}^{-1} h_{t(a)}^{-1} g_a\}$.  For instance, for the sphere nr. 1 one has 
\ba
	h_{s(41)}h_{s(31)}h_{s(21)}= \unit\,\; &\Rightarrow& \;\, h_{s(41)}=  h^{-1}_{s(21)}  h^{-1}_{s(31)} \nn\\
	& \Rightarrow& \; \, h_{t(41)} = g_{41} g_{31}^{-1} h_{t(31)} g_{31} \, g_{21}^{-1} h_{t(21)} g_{21}  g^{-1}_{41} \; .
\ea
Using the Bianchi identities for the spheres $k=1,2,3$ we can solve for $h_{t(41)}$, $h_{t(42)}$ and $h_{t(43)}$ in terms of $h_{t(21)}$, $ h_{t(32)}$, $h_{t(31)}$ and the $g$-holonomies. We are then left with the Bianchi identity for the remaining sphere. The resulting parameterization still has some gauge freedom. This can be almost completely gauge fixed  by imposing for instance the conditions $g_{41}=  g_{42} = g_{43}=\unit$. Putting everything together, we are left with a parametrization in terms of three $h$-variables and three $g$-variables. There is a  residual gauge action, which is the diagonal adjoint action, and one remaining Bianchi identity
\begin{align}\label{BianchiS3}
	g_{21}^{-1}h^{-1}_{t(21)} g_{21} h_{t(21)} \,g_{32}^{-1} h_{t(32)}^{-1} g_{32} h_{t(32)}\,  h_{t(31)} g_{31}^{-1} h_{t(31)}^{-1}g_{31} \,=\, \unit \; .
\end{align}
With this parameterization on the space of flat connections at hand, we can impose the two-handle constraints---that is the constraints that impose flatness for the holonomies around the triangles of $\triangle$. There are four triangles but, due to the Bianchi identity for the tetrahedron, we have only three independent constraints.  Furthermore, with our choice of graph, these constraints involve only $g$-holonomies, e.g. for the triangle $(124)$  we have  $g^{-1}_{41}  g_{42} g_{21}=\unit$.  With our gauge fixing we have a unique solution to these flatness constraints given by $g_a=\unit$ for all $a$.  This does also solve the remaining Bianchi identity (\ref{BianchiS3}).

Imposing the two-handle constraints leads to a first parametrization of the state space of flat holonomies on the three-sphere with a tetrahedral defect structure: It is given by (ordered) triples of  group elements $(h_{t(21)}, h_{t(32)},h_{t(31)})$ up to a global adjoint action. For the group ${\cal S}_3$ there are 49 such equivalence classes. Therefore, the Hilbert space  describing the 3d connections possesses 49 basis states.

We now would like to derive a more local parametrization, which would in particular allow us to read off directly the curvature (or magnetic charge) associated with each edge of the tetrahedral skeleton. As in the case of the genus-2 defect structure, we could hope that labeling the edges $a$ of the tetrahedron with the conjugacy classes $C_a$ of the holonomies $h_{t(a)}$ gives a one-to-one parametrization of the set of equivalence classes described above. These $C_a$ would naturally have to satisfy the coupling rules \eqref{couplingC} at each vertex of the tetrahedron. However, it turns out that the number of all such  configurations  allowed by the coupling rules is only 47. Therefore, such a parameterization would not be sufficient to capture the  entire space of 3d connections.

To investigate in more detail the failure of the parametrization by the conjugacy classes, we  reconstructed   explicitly the admissible configurations of the six holonomies $\{h_{t(a)}\}$ and their conjugacy classes $\{C_a\}$. This revealed that the  parametrization in terms of conjugacy classes fails because there is   one configuration of conjugacy classes $\{C_a\}$, which is allowed by the coupling rules but for which there does not exist a $g$-connection satisfying the two-handle constraints (case ${\sss (1)}$ in sec.~\ref{sec:curv}).  Furthermore, we have three other configurations  of conjugacy classes  $\{C_a\}$  which modulo residual gauge transformation admit two such $g$-connections  (case ${\sss (3)}$). For all other (43) configurations there exists a unique (modulo residual gauge transformation) $g$-connection satisfying the two-handle constraints (case ${\sss (2)}$). Let us illustrate this by providing explicit examples:\\

\noindent
\textsc{Case} ${\sss (1)}$: This appears for the configuration where each edge of the tetrahedron is labeled by the conjugacy class ${\cal E}$ of even elements. Thus for each vertex we have a triple $({\cal E},{\cal E},{\cal E})$ of conjugacy classes and we choose as representatives the triple $(s,s,s)$ of holonomies. The coupling condition is obviously satisfied since $s^3=\unit$. Such a choice of configuration means that all the $h_{s(a)}$- and $h_{t(a)}$-holonomies are given by the group element $s$. The question is whether we can find values for the $g_a$-holonomies, such that $(i)$ $h_{s(a)}=g_a^{-1}h_{t(a)} g_a$, and $(ii)$ the two-handle constraints hold. As explained in sec.~\ref{tDDP}, in order to satisfy $(i)$, the $g_a$-holonomies need to be of the form
\ba\label{3.17}
	g_a\,=\,    q_{\iota_{C_a}(h_t)} \cdot   z_a \cdot  q^{-1}_{\iota_{C_a}(h_s^{-1})} 
\ea
where $z_a$ belongs to the stabilizer $Z_{\cal E} =\{\unit,s,s^2\}$ and $ q_{\iota_{C_a}(h_t)}=q_{\iota_{\cal E}(s)}=\unit$. The problem arises  as we also have $q_{\iota_{C_a}(h_s^{-1})} =q_{\iota_{\cal E}(s^{-1})}=q_{\iota_{\cal E}(s^2)}=r$ since $s^2=rsr^{-1}$. From this, we can conclude that the $g_a$-holonomies are necessarily odd elements. This prevents us from satisfying $(ii)$, that is the two-handle constraints, as these require that a product of three $g$-holonomies is equal to the identity.  Thus we cannot find a $g$-connection satisfying the two-handle constraints for the configuration where all conjugacy classes are given by ${\cal E}$. It therefore makes this configuration of conjugacy classes non-admissible, despite the fact that the corresponding fusion  coefficients are non-vanishing.

~\\
\noindent
\textsc{Case} ${\sss (2)}$:  There are $43$ configurations of $h$-conjugacy classes, which are allowed by the coupling rules, and admit one and only one $g$-connection satisfying  $(i)$ $h_{s(a)}=g_a^{-1}h_{t(a)} g_a$, and $(ii)$ the two-handle constraints. This includes 39 configurations where one or more edges carry the trivial conjugacy class. There are furthermore 4 configurations for which three edges carry the conjugacy class of odd elements ${\cal O}$ and the other three edges the conjugacy class of even elements ${\cal E}$. As we are about to see, the other allowed case without a trivial conjugacy class appearing leads to a degeneracy.

\begin{figure}
	\centering
	\includegraphics[scale=0.85,valign=c]{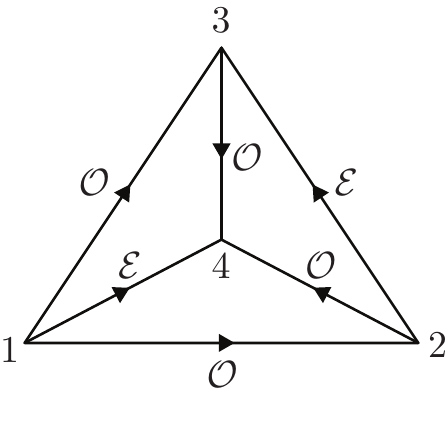} \q \q\q
	\includegraphics[scale=0.85,valign=c]{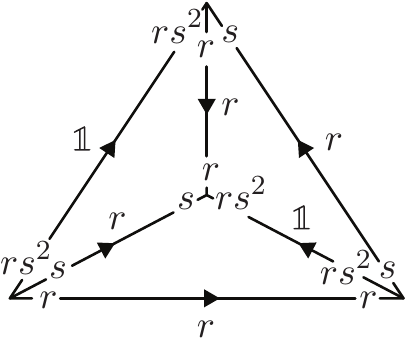}\q\q\q
	\includegraphics[scale=0.85,valign=c]{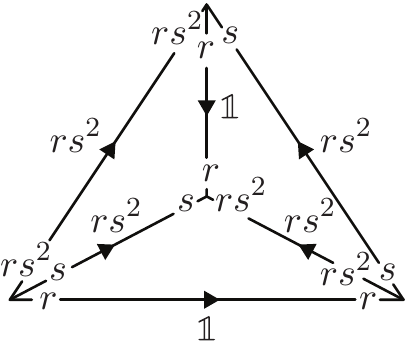}
	\caption{The left panel shows a configuration of conjugacy classes that admits two equivalence classes of solutions to the two-handle constraints. The middle and right panel show these two solutions. The $g$-holonomies are the group elements near the midpoints of the edges, the $h$-holonomies are located near the end points of the edges. The two configurations coincide in their $h$-holonomies but the $g$-holonomy configurations are different.}
		\label{fig_tetraS3}
\end{figure}

~\\
\noindent
\textsc{Case} ${\sss (3)}$: As mentioned above there are three  configurations of conjugacy classes \\$\{C_{21},C_{31},C_{41},C_{32},C_{42},C_{43}\}$ for which there are, modulo gauge transformations, two $g$-connections that satisfy the two-handle constraints. These three configurations turn out to map to each other under cyclic permutations of the vertices $(1,2,3)$ of the tetrahedron. We therefore need to discuss only one of these configurations, which we choose to be $\{
{\cal O},  {\cal O}, {\cal E},{\cal E},{\cal O},  {\cal O}   
\}$, see fig. \ref{fig_tetraS3}. A pair of non-adjacent edges carries the conjugacy class ${\cal E}$, whereas all other edges carry the conjugacy class ${\cal O}$.  As one can see from the figure, each of the three-valent vertices (being blown-up  to thrice-punctured spheres) carries the same (cyclically ordered)  triple of conjugacy classes $({\cal O},{\cal E},{\cal O})$. We choose as representative of such a triple the configuration $(r,s,rs^2)$. The stabilizer group of this triple of group elements, under the diagonal adjoint action, is given by the trivial group $\{\unit\}$.  By demanding that the $h$-holonomies on each thrice-punctured sphere (with some chosen  ordering in clockwise direction) are given by the triple $(r,s,rs^2)$, we have therefore completely fixed the gauge. 

We need now to determine the number of solutions for the $g$-connections satisfying the two-handle constraints as well as the condition $h_{s(a)}=g_a^{-1}h_{t(a)} g_a$ for each edge $a$. Considering the flatness conditions for just one triangle, e.g. the triangle $(1,3,4)$, one finds two possible solutions for the $g$-holonomies associated to the adjacent edges. These two solutions extend to two different global solutions, which we display in fig.~\ref{fig_tetraS3}. We can choose an (gauge invariant) observable with distinguishes between these two different solutions. Such an observable is given by the conjugacy class of $g_{41}^{-1}  h_{t(43)}  g_{41}h_{s(21)}$ and could be interpreted as local observable, as it involves the holonomies of one tube and the adjacent spheres only. This observable happens to distinguish the two solutions for all three degenerate configurations of $h$-conjugacy classes.

~\\
In summary, we saw that the case of a tetrahedral skeleton embedded into the three-sphere already showcases that the parametrization which attaches conjugacy classes to the edges of the skeleton is not sufficient. However, in order to obtain a complete parametrization, we  only needed to add one further gauge invariant observable. This observable combines the holonomies associated to one tube and its two adjacent punctured spheres, and in this sense can be considered to be local.  The resulting set of parameters is subject to coupling rules: Firstly, we have the coupling rules which arise from the Bianchi identities associated to the punctured spheres, which are already imposed in the fusion basis for the Heegaard surface. Secondly we have additional coupling rules which arise from the two-handle constraints. As we have seen, these constraints suppress one configuration of conjugacy classes that is a priori allowed by the Bianchi identities. Furthermore, we found that the value of the added observable is in most cases uniquely determined by the initial set of conjugacy classes and can take only two values for three configurations of this set.

\subsubsection{Three-torus}\label{3torusA}

We finally consider an example where the 3d manifold, into which the defect structure is embedded in, has a non-trivial topology. We take this 3d manifold to be a three-torus, which can be discretized as a cube with periodic boundary conditions.  Due to the periodic identification of the various elements, this lattice  has only three faces, three edges associated with the three non-contractible cycles, and one six-valent vertex. We allow for curvature defects along the edges of this one-cube lattice. We are thus interested in the space of flat connections on the three-torus with a thickening of the one-skeleton of the one-cube lattice removed.

\begin{figure}[t]
	\centering
	\includegraphics[scale=0.85,valign=c]{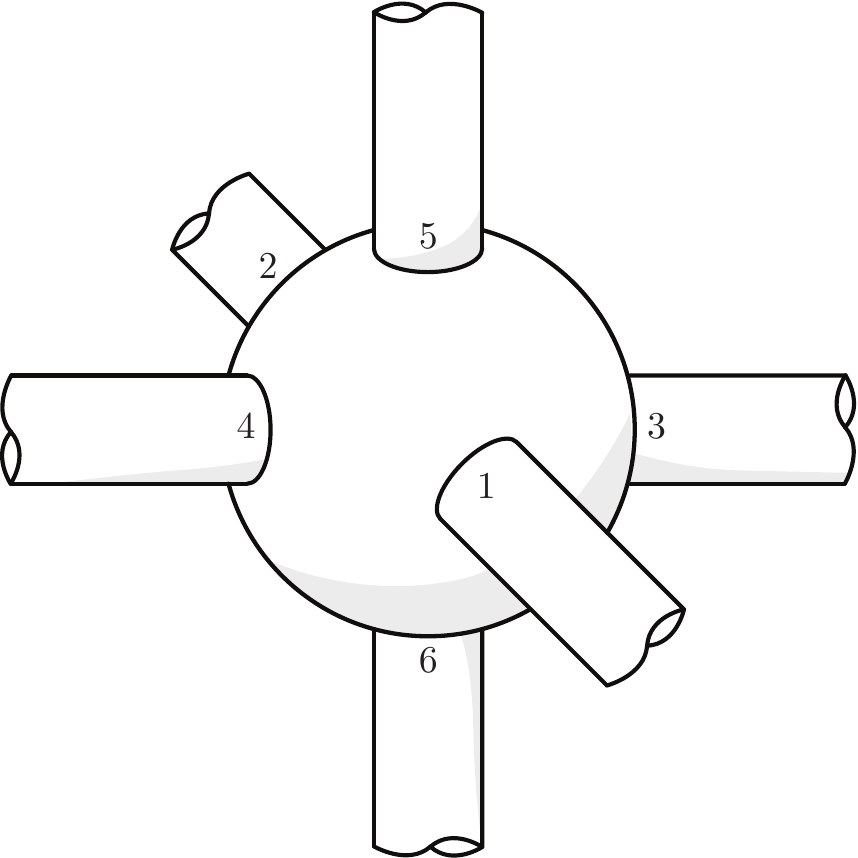} \q \q 
	\includegraphics[scale=0.85,valign=c]{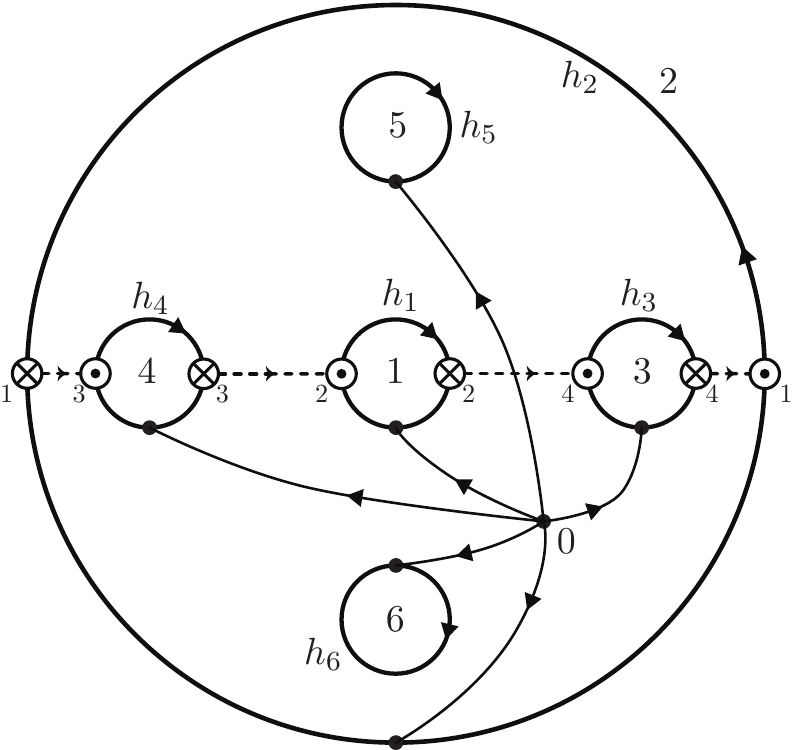}
	\caption{The three-torus can be discretized as a cube with opposite faces identified so that the discretization contains one six-valent vertex, three edges associated with the non-contractible cycles and three faces. The Heegaard surface is obtained by blowing-up the one skeleton of this discretization so that the single vertex becomes a six-times-punctured two-sphere as represented on the left panel. The edges are blown-up into cylinders that connect the punctures $1$ to $2$, $3$ to $4$ and $5$ to $6$, respectively. The same object is depicted in a different representation on the right panel as a disk with five punctures, the boundary of the disk corresponding to the sixth puncture. In this representation, the $\boldsymbol{\odot}$ of a given puncture is identified with the $\boldsymbol{\otimes}$ of the puncture at the other end of the corresponding cylinder. The dashed line represents the intersection of the Heegaard surface with the face going through the punctures $1$, 
	$2$, $3$ and $4$.}
	\label{fig_3tor1}
\end{figure}

The surface of the thickening of this one-skeleton defines a Heegaard surface for the three-torus. A convenient representation of this surface is shown in fig.~\ref{fig_3tor1}.  The sphere surrounding the one vertex with its six punctures is represented as a disk with five punctures, with the boundary of the disc defining the sixth puncture. Pairs of punctures  $(i,i+1)$ with $i=1,3,5$ are connected by tubes (or one-handles) and we have therefore a genus-3 surface.  We can find a parametrization of the space of flat connections on this surface by choosing a base node $n_0$, links from $n_0$ to the marked points $n_i$, $i=1,\ldots ,  6$ of the six punctures, links around the punctures starting and ending at $n_i$, and links along the tubes from $n_i$ to $n_{i+1}$ (see fig.~\ref{fig_3tor1}).

We associate holonomies $h_i$, $i=1,\ldots, 6$ to the (clockwise) links around the punctures and holonomies $g_{i+1,i}$ to the links from the puncture $i=1,2,3$ to the puncture $(i+1)$. Furthermore, holonomies from the base node $n_0$ to the marked points $n_i$ are gauge fixed to the identity.  Finally, the holonomies $\{h_i\}$ are not independent since
\ba
	h_{i+1}^{\phantom{-1}}=g_{i+1,i}^{\phantom{-1}} h_i^{-1} g^{-1}_{i+1,i}  \; .
\ea
We have therefore a parametrization of the space of flat connections on the genus-3 Heegaard surface by three pairs of holonomies $\{h_i,g_{i+1,i}\}_{i=1,3,5}$. This set of holonomies is subject to the Bianchi constraint associated with the six-times-punctured sphere
\ba\label{Bianchi3torus}
g_{43} h_3^{-1}g_{43}^{-1}\,g_{65}h_5^{-1}g_{65}^{-1}\,g_{21}h_1^{-1} g_{21}^{-1}\, h_3\, h_2\,h_1&=\unit  \; .
\ea
We have furthermore a left-over gauge action at the base node $n_0$, that results into a diagonal adjoint action on the set $\{h_i,g_{i+1,i}\}_{i=1,3,5}$.

We can now consider the two-handle constraints, which demand that the holonomies around the faces of the cube are flat. At this stage, some care needs to be taken in determining a path that is homotopy equivalent to the intersection of the Heegaard surface with the faces, see fig.~\ref{fig_3tor1}. It turns out that such a path does not only involve links along the tubes but needs also to wind around one of the remaining punctures. Thus, each of the three two-handle constraints involves one of the $h_i$-holonomies and can be used to fix these $h_i$-holonomies in terms of commutators of the corresponding $g$-holonomies:
\ba
	\label{3torus21}
	h_1&=&g_{65}^{-1} g_{43} g_{65} g_{43}^{-1}\nn\\
	h_3&=& g_{21} g_{65}^{-1} g_{21}^{-1} g_{65}\nn\\
	h_5&=& g_{65}^{-1}  g_{21}g_{43}   g_{21}^{-1} g_{43}^{-1} g_{65}  \; .
\ea
These relations impose automatically the Bianchi identity \eqref{Bianchi3torus}. We have thus reached a parametrization of the space of flat connections on the 3-torus with the one-skeleton of the one-cube lattice removed, which is given by  three holonomies $\{g_{21},g_{43},g_{65}\}$ modulo a diagonal adjoint action.  For the group ${\cal S}_3$ this amounts to 49 configurations (the same number as for the tetrahedron).

In the case of the three-torus, the holonomies $\{g_{i+1,i}\}$ encode global topological observables, i.e. they are associated with paths that are topologically non-trivial even without removing the blown-up one-skeleton of the discretization. Due to the small lattice, it turns out that these observables also determine completely the holonomies around the edges of the cubical discretization. Nevertheless, in order to illustrate the effects of a non-trivial underlying three-dimensional manifold, we discuss what happens if we seek a more local parametrization. 

To this end, we choose to include the conjugacy classes of the $h$-holonomies as parameters. Note that so far, we introduced the $h$-holonomies $h_1,h_3,h_5$, but we also have the set $h_2,h_4,h_6$ whose conjugacy classes coincide with those of $h_1,h_2$ and $h_3$, respectively. To gain more gauge invariant  information about the $h$-holonomies, we also consider the conjugacy classes of certain products of these $h$-holonomies:
\begin{align}
	\nn
	&C_1=C_{h_1} =C_{h_2}\q , \q C_3=C_{h_3}=C_{h_4} \q , \q C_5=C_{h_5}=C_{h_6} \q , \\ & C_{35}=C_{h_3h_5} \q ,  \q C_{62}=C_{g_{65} h_5^{-1} g_{65}^{-1}g_{21}h_1^{-1}g_{21}^{-1}} \q , \q C_{351} = C_{h_3h_5h_1}=C_{462} \; .
\end{align}
This choice defines one way to split the six-times-punctured sphere, which resulted from blowing up the one vertex of our cubical lattice, into four thrice-punctured spheres, by including virtual cylinders. The set $\{C_{35}, C_{351},C_{62}\}$ provides the conjugacy classes of the holonomies around these virtual cylinders. We already know that the tree-torus has a degenerate vacuum, that is we have non-trivial flat connections on the three-torus even if all the $h$-holonomies, which describe the (curvature) defect excitations, are trivial. Thus the   parameters $\{C_1,C_3,C_5,C_{35},C_{351},C_{62} \}$ alone fail to give an effective description for the periodically identified cubical lattice. Studying the matching of the connection configurations and these parameters in more detail, we see that 
both the cases ${\sss (1)}$ and ${\sss (3)}$ listed in sec.~\ref{sec:curv} appear.

Recall that case ${\sss (1)}$ refers to the impossibility of finding $g$-connections satisfying the two-handle constraints for certain configurations of conjugacy classes which are allowed by the coupling rules. This is evident from the form of the $h$-holonomies \eqref{3torus21} in terms of the $g$-holonomies: If a given $h$-holonomy includes a $g$-holonomy,  it must include its inverse as well. Therefore, all $h$-holonomies have to be even elements.  Thus the conjugacy class $C={\cal O}$ of odd elements cannot appear as a value for any of the parameters $\{C_1,C_3,C_5,C_{35},C_{351},C_{62} \}$. 

We also have occurences of case ${\sss (3)}$,  in which there are several equivalence classes of $g$-connections for a given configuration of conjugacy classes for the $h$-holonomies.   
As we have already mentioned, the vacuum sector is degenerate for the three-torus. There are 21 equivalence classes of $g$ connections consistent with a trivial $h$-holonomy configuration.
 This is the number of locally flat connections on the three-torus for the group ${\cal S}_3$.  Such degeneracies also appear if we have  non-trivial values for the set of $h$-conjugacy classes. Since the conjugacy class ${\cal O}$ of odd elements cannot appear for the $h$-holonomies, the only possibilities are the trivial conjugacy class ${\cal T}$ and the conjugacy class ${\cal E}$ of even elements. It turns out that the degeneracy is completely determined by how many of the six parameters   $\{C_1,C_3,C_5,C_{35},C_{351},C_{62}, \}$ take on the value ${\cal E}$. Configurations with only two such parameters are not allowed, configurations with three or four such parameters have a three-fold degeneracy, and configurations with five or six such parameters admit a unique $g$-connection.

As for the tetrahedron, it is possible to add further gauge invariant observables to the parameterization in order to resolve the degeneracies. In the case of the three-torus, we know that the $g$-connection can be non-trivial due to the non-trivial topology and it makes therefore sense to add global information about this $g$-connection. For instance, we can choose to add the conjugacy classes $C_{g_{21}}$, $C_{g_{43}}$, $C_{g_{65}}$ and $C_{g_{21}g_{43}}$, $C_{g_{43}g_{65}}$, $C_{g_{21}g_{65}}$. This does indeed get rid of all degeneracies.  This comes of course at the  price of over-parametrization.  More precisely, this adds a number of additional coupling rules that  determine which values for the $g$-conjugacy classes can appear, depending on the values for the $h$-conjugacy classes $\{C_1, C_3, C_5, C_{35}, C_{351}, C_{62}\}$.

The example we have considered here is of course extreme, as the lattice is so small that all the holonomies are determined by global parameters, namely the holonomies associated to the non-contractible cycles of the three-torus. With a more refined lattice, we expect the local parametrization to be more efficient. Of course, in the case of a non-trivial topology  for the three-manifold carrying the defects, we will always have to take into account the non-triviality of the vacuum sector.

\section{Three-manifolds with boundary: cutting and gluing \label{sec:cutglue}}

In this section, we consider a generalization of the previous construction where the three-manifold carrying the defects is allowed to have a boundary. More precisely, we discuss two kinds of cutting procedure for three-manifolds with defects: The first one is adapted to the spin network basis and is often used for the electric centre definition of entanglement entropy \cite{Donnelly2008,Donnelly2011,Casini2013, Radicevic:2014kqa,Ghosh2015, Soni2015, Delcamp:2016eya}.  The second one is rather adapted to the dual spin network basis and can be used for a magnetic centre definition of entanglement entropy \cite{Casini2013} for (3+1)d lattice gauge theories. (See \cite{Delcamp:2016eya} for a magnetic centre definition in the (2+1)d case.) We will see that the holonomy description introduced with the Heegaard surface provides an efficient way to describe such cuttings.

\subsection{Cutting}
In order to account for a boundary, we can proceed as follows. We start from a closed manifold with a given defect structure and its corresponding Heegaard surface. As a two-dimensional Riemann surface, this Heegaard surface can be obtained as a gluing of punctured two-spheres along tubes. We then cut the 3d manifold transversally through the tubes.\footnote{We could also perform the cut through a sphere. In which case, we can decompose this sphere into two spheres connected by a tube and perform the cut through this tube.} As a result of this cutting, the Heegaard surface is now decomposed into  two or more surfaces with punctures.

 As before, in order to reconstruct a 3d interpretation for these Heegaard surfaces with boundary, we need to specify which curves  are contractible in the embedding three-manifold. The holonomies along these curves are then constrained to carry a trivial holonomy. We distinguish two different cases: 

\begin{enumerate}[itemsep=0.4em,parsep=1pt,leftmargin=*]
	\item[$\circ$]
	The cut is performed parallel to a face whose boundary corresponds to a contractible loop. In this case, the face is copied for each piece resulting from the cut, so that if the pieces were glued back together, the two copies would be identified to one face. Correspondingly, the flatness constraint is imposed on each copy of the face. This situation appears when the Heegaard surface is obtained by blowing-up the dual graph $\Upsilon_1$ so that the tubes of the Heegaard surface pierce  faces of the dual triangulation. 
	\item[$\circ$]
	The cut goes through a face. In this case, we relax the corresponding flatness constraint so that it does not hold  at either of the faces resulting from the cutting. On the other hand, if we glue back the two pieces together, we do impose  the flatness constraint associated with the restored face. This situation appears when the Heegaard surface is obtained by blowing-up the one-skeleton $\triangle_1$ of the triangulation $\triangle$.
\end{enumerate}

\medskip \noindent
Following this approach, we can define the Hilbert space of flat connections for a three-manifold with defects and boundary. We do so by first defining the Hilbert space of flat connections on the associated punctured Heegaard surface and then impose flatness constraints at every face fully included into the triangulation. We discuss explicit examples below. In general, such a cutting procedure is best defined as the dual procedure of the corresponding gluing. We introduced how to glue two dimensional surfaces so as to define  the tube algebra in sec.~\ref{sec:tube}. We can now discuss a three-dimensional version of this gluing.

\subsection{Gluing} 

Let us define how to glue pieces of three-manifolds, and the corresponding Hilbert spaces of flat connections, together. As far as the gluing of punctured 2d surfaces is concerned, the basic idea has been laid out in \cite{Delcamp:2016eya} and was recalled in sec.~\ref{sec:tube}. Since we are able to represent flat connections on 3d manifolds with defects and with boundary as flat connections on punctured 2d surfaces, we just need to add the two-handle flatness constraints to the discussion.

Let ${\cal H}_A$ and ${\cal H}_B$ be two Hilbert spaces of flat connections on the punctured surfaces $\Sigma_A$ and $\Sigma_B$, respectively. These punctured surfaces represent two pieces of a Heegaard surface obtained as the blow-up of two pieces of a triangulation so that they can be glued to each other. It is understood that the states in ${\cal H}_A$ and ${\cal H}_B$ satisfy the flatness constraints associated with the faces that are fully included in the triangulations. Performing the gluing of the two pieces of triangulation implies the gluing of the two pieces of the Heegaard surface. As before, each puncture of the two-dimensional surfaces is assigned a marked point. Pairs of these marked points need to be properly identified upon gluing. Furthermore, each piece of the Heegaard surface is equipped with an embedded graph such that the marked points serve as one-valent nodes. Identifying the marked points then ensures that the embedded graphs are properly joined upon gluing of the corresponding surfaces. 

Given two states $\psi_A \in {\cal H}_A$ and $\psi_B \in {\cal H}_B$, we defined in sec.~\ref{sec:tube} the $\star$-product \eqref{starProd}: We first form a new state $\psi_A \cdot \psi_B$ on the graph $\Gamma_{A \cup B}$ obtained by connecting the graphs $\Gamma_A$ and $\Gamma_B$ embedded on $\Sigma_A$ and $\Sigma_B$, respectively. We then impose constraints at the new nodes and faces emerging from the gluing. We distinguish three types of constraints when dealing with Heegaard surfaces representing three-manifolds, the first two being common to the case of  gluing  2d surfaces:
\begin{enumerate}[itemsep=0.4em,parsep=1pt,leftmargin=*]
	\item[$\circ$] The Gau\ss~constraints demanding gauge invariance at the two-valent nodes that arise from identifying the marked points of two punctures with each other.
	\item[$\circ$] The flatness constraints that arise from newly formed faces of the graph embedded on the Heegaard surface.
	\item[$\circ$] The two-handle flatness constraints associated with the newly formed triangles of the glued 3d triangulation.
\end{enumerate}

\subsection{Cutting in the spin network basis}

Let us assume the Heegaard surface is chosen such that it is adjusted to constructing the spin network basis. We decompose the Heegaard surface into punctured spheres surrounding the nodes of the 3d dual graph and into tubes surrounding the links of the 3d dual graph. Cutting transversally through the tubes, we cut along the triangles of the triangulation. Therefore, the boundary only contains full triangles. Accordingly, we do impose the flatness constraints associated with these boundary triangles as well as with all bulk triangles.

In the case without boundary, we can associate a fusion basis to such a decomposition of the Heegaard surface adjusted to the dual graph.  This fusion basis associates labels $(C_t,R_t)$ to each tube of the surface decomposition---that is for each triangle of the triangulation. There are possibly additional intertwiner labels for the spheres (resulting from the decomposition of $n>3$ punctured spheres into thrice-punctured spheres).  In case the tubes are open, i.e. are associated to a boundary triangle $t$, we have an additional vector space index $N_t$ associated with the representation space $V_{C_t,R_t}$. 

Imposing the flatness constraints associated with the triangles force again all conjugacy class labels $C_t$ to be trivial. Therefore, the $R_t$ labels stand for representations of the full group and the vector space indices $N_t$ reduce to vector space indices associated with $V_{R_t}$. We obtain a spin network basis with open edges, that can carry \emph{electric} indices. These edges ending at the boundary can be understood to carry a Gau\ss~constraint violation or electric charge at their end. This structure corresponds to the `electrically' extended Hilbert space defined in \cite{Donnelly2008,Donnelly2011}. This choice is associated to a definition of entanglement entropy with an \emph{electric centre} \cite{Donnelly2008,Donnelly2011,Casini2013, Ghosh2015, Soni2015, Delcamp:2016eya}.

\subsection{Cutting in the curvature basis and extended quantum triple algebra}

Let us now assume the Heegaard surface is chosen such that it is adjusted to constructing the curvature basis. We only allow cuts through the tubes surrounding edges of the triangulation (or lattice) so that we cut through the triangles (or faces of the lattice). Only the flatness constraints for triangles and faces that are fully included in the triangulation with boundary are imposed.  The Hilbert space, for a triangulation with a boundary resulting from such a cut, is defined from the Hilbert space of flat connections on the associated punctured surface by imposing the two-handle constraints for all bulk triangles (or more generally bulk faces).

Let us consider a concrete example, namely the three-torus  with the one-cube lattice sliced open along one plane.  In the notation of sec.~\ref{3torusA}, we assume this slicing to cut through the tube between the punctures  $i=5$ and $i=6$. The resulting 2d surface is of genus two with two punctures. We keep most of the notation of sec.~\ref{3torusA}, but now have to take into account that $h_5$ and $h_6$ are, a priori, two independent holonomies around the punctures $i=5$ and $i=6$, respectively. Furthermore, we replace the holonomy $g_{65}$ with the holonomies $g_5$ and $g_6$ that go from the marked point of the puncture $5$ and $6$, respectively, to the marked point at the other end of the `half-tube'. Hence the space of flat connections on the 2d surface is parametrized by the holonomies $\{h_1,h_3,h_5,h_6\}$ and $\{g_{21},g_{43}, g_5, g_6\}$.  These are subject to the Bianchi identity
\ba\label{5.1}
g_{43}h_3^{-1} g_{43}^{-1} h_6 g_{21}h_1g_{21}^{-1} h_3 h_5 h_1 \,=\, \unitG 
\ea
which can be solved for $h_5$ in terms of the other holonomies.
Furthermore, there is a residual gauge action that acts as a diagonal adjoint action on all the holonomies, except on $g_5$ and $g_6$ for which it amounts to a right action. This can be used to obtain a complete gauge fixing, i.e. by demanding that $g_5=\unitG$. In summary, the space of flat connections on our genus-two surface with two punctures is parametrized by the holonomies $\{h_1,h_3,h_6\}$ and $\{g_{21},g_{43},g_6\}$ without any further constraints to impose or gauge equivalences to take into account.

But now we have to consider the flatness constraints associated with the faces of the one-cube lattice. Since only the face spanned by the tubes $(12)$ and $(34)$ is intact, we have a single two-handle constraint
\begin{equation}
	\label{resface}
	h_6=g_{43}g_{21}g_{43}^{-1}g_{21}^{-1} 
\end{equation}
that allows to fix $h_6$. Thus we obtain a space of flat connection for the one-cube lattice periodically identified along two directions parametrized by five holonomies $\{h_1,h_3\}$ and $\{g_{21},g_{43},g_6\}$. We can glue back the slice open three-torus to a three-torus, which in particular requires imposing the two flatness constraints associated with the two new faces. This gives back the configuration space of the three-torus with defects. 

Another possibility is to glue two slice open three-tori together---which topologically gives again a sliced open three-torus. In the case that we impose vanishing curvature defects on the edges, the two pieces that we glue and the piece that we end up with all have the same topology and therefore carry equivalent Hilbert spaces of flat connections.  This allows to define an algebra referred to as \emph{quantum triple} in \cite{Wan:2014woa, Delcamp:2017pcw}. Here we consider a slight extension of this algebra by  allowing also non-trivial curvature along the edge transversal to the gluing plane (i.e. we have non-trivial $h_5=h_6^{-1}$).  

Let us compute explicitly this \emph {extended quantum triple algebra} resulting from such gluing. Setting  $h_1=h_3=\unitG$, the free parameters for the slice open three-torus are given by $(g_{21},g_{43},g_6)$. These holonomies do determine also $h_5$ and $h_6$ via \eqref{resface} and \eqref{5.1},  respectively. We now consider another slice open three-torus with parameters $(g_{2'1'},g_{4'3'},g_{6'})$ which  we glue to the first one by identifying its puncture nr. 6' to the puncture nr. 5 of the previous one. After gluing the graphs on which the holonomies are based together, we have to impose  the flatness constraints for the 2d surface, the two-handle constraints and the gauge invariance for the new two-valent vertices. More precisely, we need to consider the flatness constraints for the new tube going from puncture 5 to puncture $6'$ and the gauge invariance for the two-valent node  on this tube. Additionally, there are two two-handle constraints to impose. Renaming $g=g^{-1}_6,g_{21}=h_1,g_{34}=h_2$, and likewise for the variables with primed indices, and introducing the notation $|g,h_1,h_2\rangle$ and $|g',h'_1,h'_2\rangle$ for (basis) states on the slice open three-torus, the algebra resulting from the gluing is a generalization of \eqref{DDproduct} in sec.~\ref{sec:tube}:
\ba\label{5.3}
|g',h'_1,h'_2\rangle\, \star \,|g,h_1,h_2 \rangle \,=\, \delta(h'_1,g' h_1 (g')^{-1})\, \, \delta(h'_2,g' h_2 (g')^{-1})\, \, |g'g, \,h'_1,\, h'_2\rangle \; .
\ea
This reproduces the multiplication for the quantum triple algebra as defined in \cite{Delcamp:2017pcw} at the difference that we allow for a defect parallel to the direction of the gluing so that we do {\it not} impose the flatness conditions $[h_2,h_1]=\unitG$ and $[h_2' ,h_1']=\unitG$. 

Similarly to the tube algebra, we can ask for the irreducible modules of this algebra. These irreducible modules can then be identified with the charge sectors.  More specifically, we are interested in the irreducible representations of this tube algebra. These can be constructed as a generalization of the Drinfel'd double representations, discussed in app.~\ref{app_drinfeld}, and the quantum triple representations introduced in \cite{Delcamp:2017pcw}. As explained in more detail in app.~\ref{RepQT}, the representations are labeled by a pair $(T,R)$. Here $T$ is a  triple of conjugacy classes $T=(C_1, C_2, C_3, d)$ together with a degeneracy variable $d$. We only admit triples for which there are group elements  $h_1 \in C_1, h_2 \in C_2$ and $h_3 \in C_3$ that satisfy
\begin{align}
	\label{5.4}
	h_3= h_2 h_1 h_2^{-1}h_1^{-1}  \; .
\end{align}
The label $d$ differentiates between potentially multiple orbits of the adjoint action in $ {\cal C} \subset G\times G\times G$, where ${\cal C}$ is the subset satisfying \eqref{5.4}. The label $R$ in stands for an irreducible representation of the stabilizer group of some representative element in the orbit associated to $T$. As for the Drinfel'd double, we can use  the irreducible representations to define a basis for the state space defined on the slice open torus. This basis would have labels $(T,R; a,m; b,n)$ where $a,b$ is a pair of labels for the elements in the orbit associated to $T$ and $m,n$ are vector space indices for the representation $R$. The labels for this basis include the conjugacy class of the defect in addition to the conjugacy classes of the two global holonomies of the slice open three-torus.  Thus the labeling of defect excitations with conjugacy classes, resulting from choosing a 2d fusion basis adjusted to the one-skeleton of the triangulation,  is also justified if we consider 3d manifolds and their gluing. 

\medskip \noindent
As a side result, we note that the extended quantum triple algebra can be easily generalized.
We have so far three examples of algebras which result from gluing flat connection configurations on manifolds with boundary. The manifolds have the topology $\left[0,1\right] \times \Sigma$ where $\Sigma$ is the circle $\mathbb{S}_1$ for the Drinfel'd double algebra, the two-torus $\mathbb{T}_2$ for the quantum triple and finally the once-punctured two-torus for the extended quantum triple discussed in this section. These examples suggest that we can generalize this approach to any 2d surface $\Sigma$\footnote{The algebra can be also generalized to higher-dimensional $\Sigma$ along the same lines.} with or without punctures. Considering only gauge invariant configurations, that is we do not allow for open edges on $\Sigma$, the space of flat connection on $\Sigma$ can be parametrized by a set of $n$ holonomies  
$
\{h_1, \ldots, h_n\}/G
$
quotiented by the adjoint action and possibly with a constraint ${\cal C}$, which is invariant under the adjoint action, imposed.  The space of flat connections on $\left[0,1\right] \times \Sigma$ is then given by $\{g,h_1, \ldots, h_n\}/G$ where the $h$-holonomies have to satisfy ${\cal C}$. The algebra is given by
\begin{align}
	|g',h'_1, \ldots, h'_n\rangle \,\star \, |g,h_1,\ldots, h_n\rangle \,=\,   \bigg( \prod_{i=1}^n \delta(h'_i, g' h_i(g')^{-1}) \bigg) 
	\, |g'g,h'_1, \ldots, h'_n\rangle \; ,
\end{align}
where we assume that the constraint ${\cal C}$ is satisfied.
The irreducible representations of this algebra are labeled by an orbit $T$ of the adjoint action in the constrained submanifold described by ${\cal C}$ in $G^n$ and an irreducible representation $R$ of the stabilizer group of a representative point in the given orbit $T$ under the adjoint action. The representation matrices can be defined in analogy to the case where $\Sigma$ is a punctured torus, see app.~\ref{app_puncS}.

\newpage

\section{Discussion}

In this work, we have introduced a novel way to construct gauge invariant bases for (3+1)d lattice gauge theories.  In particular, we have reconstructed the known spin network basis but found also a new basis referred to as the dual spin network basis. The latter provides a local description of magnetic excitations for $BF$ topological phases  and is  particularly suited for the new representations\footnote{Although one uses a Lie group $\SU(2)$, this group is equipped with a discrete measure, which makes this case similar to a discrete group. We therefore expect that possible measure theoretic questions can be resolved.} of loop quantum gravity based on the $BF$ vacuum \cite{DGflux,DGfluxC, DGfluxQ,Lewandowski:2015xqa}. The duality between these two bases for (3+1)d lattice gauge theories was first exposed for the quantum group case in \cite{Dittrich:2017nmq}. Moreover, it is reminiscent of a similar duality in (2+1)d, discussed in  \cite{Delcamp:2018sef}.\footnote{There, two quantization schemes are presented which correspond to different orderings of implementation of the constraints of the theory, namely the Gau{\ss} constraint and the flatness constraint. The first constraint to be implemented, i.e. the \emph{kinematical one}, dictates a choice of representation and \emph{a fortiori} basis for the kinematical Hilbert space. When the flatness constraint is the kinematical one, it is natural to work with a basis labeled by group variables instead of the usual spin network basis.} Furthermore, in the context of the 3d Turaev-Viro model and the Ponzano-Regge model, a related duality, based on a change of variables for the corresponding partition functions, has been described in \cite{Barrett:2004im} and \cite{Freidel:2005bb}, respectively.

In order to construct these bases, we have employed a lifting technique that uses the representation of three-dimensional manifolds on so-called Heegaard surfaces. States and Hilbert spaces of (2+1)d topological theories are mapped to states and Hilbert spaces of a (3+1)d theory with certain types of line defects.  We have also used this lifting technique to describe state spaces for three-dimensional manifolds or lattices with boundaries and their gluing. We encountered two types of boundaries: the well-known electric boundary \cite{Donnelly2011, Donnelly2014}, where one cuts through the links of the spin network, and an alternative magnetic
boundary, where one cuts through the edges of the defect structure, e.g. a triangulation. These different kinds of boundary lead to different notions of entanglement entropy for gauge theories, namely the electric centre and magnetic centre definition, respectively \cite{Casini2013,Radicevic:2014kqa}. This has been already shown in 2+1 dimensions \cite{Delcamp:2016eya}. So far there has not been any explicit definition for the magnetic centre version in 3+1 dimensions for non-abelian groups, but we expect that one can find such a definition following \cite{Delcamp:2016eya} and the gluing procedure discussed here. Note that the gluing of the magnetic boundary requires to employ a gauge averaging, as for the electric boundary, but also to enforce flatness and two-handle constraints. We expect these constraints to lead to additional contributions for the magnetic centre definition of entanglement entropy. Furthermore, we showed how we can use the gluing with magnetic boundary to derive a generalization of the so-called quantum triple algebra \cite{Delcamp:2017pcw}, which similarly to the quantum double or tube algebra in (2+1)d, can be used to classify excitations for (3+1)d phases. 

Let us comment more on the two different types of bases constructed using the lifting technique. The spin network basis diagonalizes certain combinations of electric flux operators, whereas the dual spin network basis diagonalize (magnetic) Wilson loop operators.  In the context of gauge models of topological phases, the dual spin network basis characterize magnetic excitations in a local\footnote{A much more global characterization results from the holonomy parametrization of sec.~\ref{holp}, which is equivalent to a gauge fixing  along a spanning tree of the dual graph. } manner and thus defines an {\it excitation basis}. The construction of the spin network basis via the lifting procedure is straightforward. The states of the spin network basis are uniquely characterized by a labeling of the graph dual to the defect structure (e.g. given by a triangulation) with irreducible representations of the group associated to links and intertwiners associated to the nodes. We would expect the dual spin network to be very similar to the spin network basis. Naively, we would replace the graph dual to the triangulation with the one-skeleton of the triangulation, and the irreducible representations of the group with its conjugacy classes. However, it turns out to be more complicated. Indeed, to each admissible set of conjugacy classes, there might correspond a unique (gauge invariant) configuration, several configurations, or no configurations at all.  Which case occurs depends on the number of  solutions to the two-handle constraints modulo gauge equivalence. We have seen that in two very simple cases the mapping between admissible labelings with conjugacy classes and gauge invariant configurations is unique, but this is generally not the case.

Due to the degeneracy of locally flat configurations on manifolds with non-trivial topology, we have to expect a mismatch at least in this case. Indeed, we find that all three types of matching and non-matching do occur for the three-torus. But we have also found an example, namely the tetrahedral lattice embedded in the three-sphere, in which all three types occur in spite of the topology of the manifold being trivial. This mismatch can be improved by adding further (Wilson-loop) observables. Furthermore, the case of no solutions for the two-handle constraints suggests the existence of---possibly non-local---coupling rules. A complete resolution of this issue requires the understanding of the intricate interplay between group structure and the topology of the lattice.

It turns out that the construction of the dual spin network basis is very similar to the task of finding a complete (and not over-complete) set of Wilson loops that characterize the  holonomy configurations on a given lattice. The main problem here are the so-called \emph{Mandelstam identities} which relate the values of different Wilson loops \cite{Mandelstam:1968hz}.  Certain progress has been achieved to solve these {Mandelstam identities} and to determine a set of independent parameters, but only for particular (regular) lattices and for specific groups (e.g. $\SU(2)$) modulo certain discrete ambiguities \cite{Loll:1991mh,Loll:1992fk, Watson:1993zr}. The local counting argument in \cite{Watson:1993zr} indicates that, in the case of $\SU(2)$, the admissible labelings of the triangulation with conjugacy classes does indeed lead to the correct number of continuous parameters, but additional discrete parameters might be required.\footnote{We also expect this to hold only for an open neighbourhood in the set of gauge invariant configurations and not for its boundaries or conical singularities.} This is also consistent with the fact that the quantum group $\SU(2)_{\rm k}$ does admit a dual spin network basis in which these labelings are indeed necessary and sufficient \cite{Dittrich:2017nmq}. It would be interesting to revisit this work addressing the Mandelstam identities, which is based on more local considerations, in the light of the technique presented here, which provides a more global view.  The resolution of these issues would for instance allow for a new understanding of the Yang-Mills (lattice) dynamics in terms of Wilson loops \cite{Makeenko:1980vm,Loll:1992fk}.

Finally, we want to emphasize the main differences between the (2+1)d and the (3+1)d cases.  In (2+1)d, we have the spin network basis as well as the fusion basis. The fusion basis may be seen as being dual to the spin network basis, but it encodes both electric and magnetic excitations. Furthermore, in the case of non-abelian groups, the fusion basis has an inherent hierarchical structure and describes an entire fusion scheme, starting with the excitations at the lattice level and extending to the most coarse-grained one.  The algebraic input for the fusion basis are representations of the Drinfel'd double of the group. It turns out that we obtain a very similar structure when working with (root of unity) quantum groups. In the language of topological phases or string net models \cite{Lan2013}, it amounts to working with a modular fusion category instead of a pre-modular one. However, in 3+1 dimensions, the cases of  pre-modular and modular fusion categories are quite different. This starts with the fact that pre-modular fusion categories lead to a degenerate vacuum\footnote{This is essentially the statement that $BF$ theory is non-trivial as a topological theory.} whereas modular fusion categories lead to a unique vacuum.  The case of the modular fusion category appears to be much simpler, in the sense that a systematic definition of a dual spin network basis can be obtained. On top of that we have a self-duality which follows from the fact that both bases are labeled by objects of the modular fusion category \cite{Dittrich:2017nmq}.

These differences can be understood in the light of the fact that the (2+1)d case is governed by the Drinfel'd double (or more precisely the Drinfel'd center), which for both modular and pre-modular fusion categories, gives a modular fusion category. The lifting procedure shows that for the (3+1)d case, the two-handle constraints force us to work with `half' the Drinfel'd double. This is straightforward in the case where the original fusion category was modular but rather non-symmetric in the case of a non-modular one.  This also suggests two interesting generalizations, which can lead to topological theories with new types of defects. The first is to  replace the magnetic-type two-handle constraints  with electric type two-handle constraints. The second is to not impose the two-handle constraints at all and in this way  allow for excitations along the triangulation as well as excitations along the dual graph. 

We hope that the different bases constructed here will help to construct suitable new tensor network coarse-graining schemes for lattice gauge theories and spin foam models \cite{Dittrich:2011zh,Dittrich:2014mxa,Delcamp:2016dqo,Dittrich:2016tys,Dittrich:2014ala}. Since the fusion of purely magnetic charges can lead to electric charge components, the spin network basis is not stable under coarse graining.  It rather requires some kind of extension, see \cite{Charles:2016xwc} for proposals. We expect that this holds also for the dual spin network basis, as it also describes only magnetic excitations. However, the construction we presented here is based on lifting the (2+1)d fusion basis to (3+1)d, the fusion basis being stable under coarse graining. We therefore believe that the strategy presented in this paper can be used in order to construct suitable extensions of the (3+1)d spin network basis and dual spin network basis so as to be stable under coarse-graining.

\vspace{1cm}

\begin{center}
\textbf{Acknowledgements}
\end{center}
The authors would like to thank Aldo Riello for helpful discussions and comments. 
CD is supported by an NSERC grant awarded to BD. 
This research was supported in part by Perimeter Institute for Theoretical Physics.
Research at Perimeter Institute is supported by the Government of Canada through the Department of Innovation, Science and Economic Development Canada and by the Province of Ontario through the Ministry of Research, Innovation and Science.

\newpage

\appendix
\section{Drinfel'd double \label{app_drinfeld}}
The Drinfel'd double of a group $G$ is an example of quasi-triangular Hopf algebra. This appendix recalls some important features of this algebraic structure that are useful for our construction. More details can be found in e.g. \cite{drinfel1988, Koornwinder1996, Koornwinder1998, Koornwinder1999} as well as in \cite{DDR1} where numerous formulas were derived.

 As a vector space, the Drinfel'd double of a group $G$ is isomorphic to
\be
	\mD(G) \simeq \mathbb{C}[G] \otimes \mathcal{F}(G)
\ee
where $\mathbb{C}[G]$ is the group ring and $\mathcal{F}(G)$ is the abelian algebra of linear functions on $G$. A basis for $\mD(G)$ is therefore provided by $\{g \smo \delta_h\}_{g,h \in \mG}$ where $\delta_h({\sss \bullet}) \equiv \delta(h,{\sss \bullet}) \equiv \delta_{h,{\sss \bullet}}$ is the delta function peaked on $h$. As a Hopf algebra, the Drinfel'd double comes equipped with a multiplication rule
\begin{align}
	\star \;\; : \;\; \mD(G) \otimes \mD(G) \;\;\;& \longrightarrow \; \mD(G)\\
	\big( g_1 \smo \delta_{h_1}, g_2 \smo \delta_{h_2} \big) \;&\longmapsto \; (g_1 \smo \delta_{h_1}) \star (g_2 \smo \delta_{h_2}) = \delta_{h_1,g_1 h_2 g_1^{-1}} (g_1g_2 \smo \delta_{h_1})
\end{align}
and the corresponding unit element is given by $\mathbbm{1}_{\mD(G)}=\sum_{h \in G}\mathbbm{1}_G \smo \delta_h$, where $\mathbbm{1}_G$ is the unit element of the group $G$. It is also equipped with a compatible comutliplication
\begin{align}
	\Delta \;\; : \;\; \mD(G) \;& \longrightarrow \; \mD(G) \otimes \mD(G)\\
	g \smo \delta_h \;&\longmapsto \; \Delta(g \smo \delta_h) = \sum_{x,y \in G \atop xy = h}(g \smo \delta_x) \otimes (g \smo \delta_y) 
\end{align}
and the corresponding counit is defined by $\epsilon(g \smo \delta_h) = \delta_{h,\mathbbm{1}_G}$.
The antipode, which can be thought as a generalization of the inverse element in the context of Hopf algebras, is defined via the relation
\be
		\star \circ (\rm{id} \otimes S)\circ \Delta = \star \circ (S \otimes {\rm id}) \circ \Delta = \mathbbm{1} \circ \epsilon \; , 
\ee
and its explicit expression reads
\be
	S(g \smo \delta_h) = g^{-1} \otimes \delta_{g^{-1}h^{-1}g} \; .
\ee
The irreducible representations $\{\rho=(C,R)\}$ of $\mD(G)$ are labeled by a conjugacy class $C$ and an irreducible representation $R$ of the centralizer $Z_C$ of $C$. The elements of $C$ are denoted $c_a$ and its representative $c_1$. The elements of the quotient $Q_C \simeq G / Z_C$ are denoted by $q_a$ such that $c_a = q_a c_1 q_a^{-1}$. Finally, the matrix elements of the Drinfel'd double element $g \smo \delta_h$ in the representation $(C,R)$ are given by
\be
	\label{defReps}
	D^{C,R}_{am,bn}(g \smo \delta_h) = \delta(h,c_a)\delta(c_a,gc_bg^{-1})D^R_{mn}(q_{a}^{-1}gq_b)
\ee
where $m$ and $n$ are the magnetic indices associated to the representation $R$ of $Z_C$. Thereafter, we make use of the more compact notation $D^{\rho}_{MN} \equiv D^{C,R}_{am,bn}$ such that $\rho \equiv C,R$, $M \equiv am$ and $N \equiv bn$. By tracing over the magnetic indices, we obtain the defining equation for the characters
\be
	\chi^{\rho}(g \smo \delta_h) = 						\delta(gh,hg)\Theta_C(h)\chi^R(q^{-1}_{\iota_C(h)}gq_{\iota_C(h)})
\ee
where $\Theta_C( {\sss \bullet})$ is the characteristic function of $C$ and $\iota_C({\sss \bullet})$ is the labeling function defined such that $\iota_C(c_a) = a$. 
The set of irreducible representations is complete and orthogonal with the orthogonality being provided by
\be
	\frac{1}{|G|}\sum_{g,h \in G}D^{\rho_1}_{M_1N_1}(g \smo \delta_h)\overline{D^{\rho_2}_{M_2N_2}(g \smo \delta_h )}
	= \frac{\delta_{\rho_1,\rho_2}}{d_{\rho_1}}\delta_{M_1,M_2}\delta_{N_1,N_2} 
\ee
where $d_{\rho} = d_{C,R} = d_R . |C|$ denotes the dimension of the representation $\rho$. Furthermore, the matrix elements of the representation $\rho^\ast$ dual to $\rho$ are obtained via the antipode as follows
\be
	D^{\rho^\ast}_{MN}(g \smo \delta_h) = D^\rho_{NM}(S(g \smo \delta_h)) = D^\rho_{NM}(g^{-1} \smo \delta_{g^{-1}h^{-1}g})\; .
\ee
We will be careful in distinguishing the last expression with the one for the complex conjugate of the matrix elements
\be
	\overline{D^{\rho}_{MN}(g \smo \delta_h)} = D^\rho_{NM}(g^{-1} \smo \delta_{g^{-1}hg})\; .
\ee
Thanks to the comultiplication, tensor product of representations can be constucted:
\be
	(D^{\rho_1}\otimes D^{\rho_2})(\Delta(g \smo \delta_h)) = \sum_{x,y\in G \atop xy = h}(D^{\rho_1}\otimes D^{\rho_2})((g \smo \delta_x)\otimes (g \smo \delta_y)) \;.
\ee
Such tensor product can then be decomposed into irreducible representations according to the fusion rules $N^{\rho_3}_{\rho_1 \rho_2}$ {\it i.e.}
\be
	\rho_1 \otimes \rho_2 = \bigoplus_{\rho_3} N^{\rho_3}_{\rho_1 \rho_2}\, \rho_3 \;.
\ee
The fact that the comultiplication is an algebra homomorphism further implies the existence of a unitary map $\mathcal{C}^{\rho_1 \rho_2}: \bigoplus_{\rho_3 \in \rho_1 \otimes \rho_2}V_{\rho_3}\rightarrow V_{\rho_1} \otimes V_{\rho_2}$ which satisfies
\be
	D^{\rho_1}_{M_1N_1}\otimes D^{\rho_2}_{M_2N_2}(\Delta(g \smo \delta_h)) = \sum_{\rho_3}\sum_{M_3N_3}
	\mathcal{C}^{\rho_1\rho_2\rho_3}_{M_1M_2M_3}\, D^{\rho_3}_{M_3N_3}(g \smo \delta_h)\,\overline{\mathcal{C}^{\rho_1\rho_2\rho_3}_{N_1N_2N_3}}\; .
\ee
We refer to such maps as Clebsch-Gordan coefficients by analogy with the group case. In the main text, we make use of the following intertwiners refer to as $3 \rho M$-symbols
\be
	\Big( {}^{\, \rho_1 \; \rho_2 \; \rho_3}_{N_1N_2N_3}\Big)
	:= \frac{1}{\sqrt{d_{\rho_3}}}\mathcal{C}^{\rho_1\rho_2\rho_3^\ast}_{N_1N_2N_3}\; .
\ee
From the unitarity of $\mathcal{C}^{\rho_1 \rho_2}$, it follows the orthogonality relation
\be
	\sum_{N_1,N_2}\Big({}^{\, \rho_1 \; \rho_2 \; \rho}_{N_1 N_2 N}\Big) \overline{\Big({}^{\, \rho_1 \; \rho_2 \; \rho'}_{N_1 N_2 N'}\Big)} = \frac{1}{d_{\rho}} \delta_{\rho,\rho'}\delta_{N,N'},
	\label{CGortho}
\ee
as well as the completeness relation
\be
	\sum_{\rho}\sum_N d_{\rho} \Big({}^{\, \rho_1 \; \rho_2 \; \rho}_{M_1 M_2 N}\Big) \overline{\Big({}^{\, \rho_1 \; \rho_2 \; \rho}_{N_1 N_2 N}\Big)} = \delta_{M_1,N_1}\delta_{M_2,N_2} \;.
	\label{CGcomplete}
\ee

\section{S-matrix for the punctured torus}\label{app_puncS} 

The construction of a  fusion basis  involves a choice of pant decomposition for the surface $\Sigma$. Different choices can  be related via so-called \emph{$S$-transforms} and \emph{$F$-moves}. The $F$-moves relate the two different ways of obtaining a four-punctured sphere by gluing two thrice-punctured spheres. Such a map is obtained by defining the $F$-symbols for the Drinfel'd double. These are simply defined as a contraction of four Clebsch-Gordan coefficients. On the other hand, the $S$-transform provides the equivalence between the two possible bases on a punctured torus $\mathbb{T}^1_2$. In this appendix, we derive the expression for the matrix elements of such $S$-transform.

 The defining formulas for these two bases, labeled by $A$ and $B$, respectively, are
\begin{align}
	|\rho_1\rho_2,N_2 \ra^A_{\mathbb{T}^1_2} &= 
	\frac{1}{|G|}\sum_{g,h}\sum_{M_1,N_1}\sqrt{d_{\rho_1}}D^{\rho_1}_{M_1N_1}(g \smo \delta_h)
	\Big({}^{\, \rho_1^\ast \; \rho_1 \; \rho_2^\ast}_{M_1N_1N_2}\Big)\; |g,h\ra_{\mathbb{T}^1_2} \\
	|\rho_1\rho_2,N_2 \ra^B_{\mathbb{T}^1_2} &= 
	\frac{1}{|G|}\sum_{g,h}\sum_{M_1,N_1}\sqrt{d_{\rho_1}}D^{\rho_1}_{M_1N_1}(g^{-1}hg \smo \delta_{g^{-1}})
	\Big({}^{\, \rho_1^\ast \; \rho_1 \; \rho_2^\ast}_{M_1N_1N_2}\Big)\; |g,h\ra_{\mathbb{T}^1_2}
\end{align}
which are obtained by performing a generalized Fourier transform of the states $| g,h\ra_{\mathbb{T}^1_2}$ and $|h^{-1},h^{-1}gh \ra_{\mathbb{T}_2^1}$, respectively.
We are looking for coefficients $S^{\rho_2}_{\rho_1\widetilde{\rho}_1}$ satisfying
\be
	|\rho_1\rho_2,N_2 \ra_{\mathbb{T}^1_2}^A = \sum_{\widetilde{\rho}_1}S^{\rho_2}_{\rho_1\widetilde{\rho}_1}
	|\wro_1\rho_2,N_2 \ra_{\mathbb{T}^1_2}^B \; .
\ee
\begin{figure}[t]
	\centering
	\label{fig_punctori}
	\includegraphics[scale=0.85,valign=c]{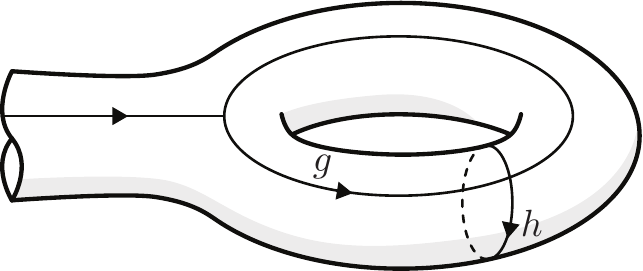} \q $\stackrel{S}{\longleftrightarrow}$ \q
	\includegraphics[scale=0.85,valign=c]{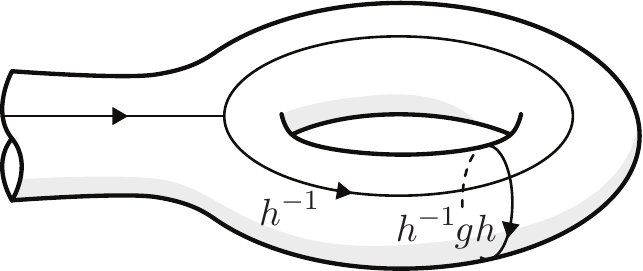}
	\caption{Graphical depiction of the $S$-transform on the punctured two-torus.}
\end{figure}
Starting from the defining formula of $| \rho_1 \rho_2,N_2 \ra^A_{\mathbb{T}_2^1}$, we can obtain a relation in terms of $| \rho_1 \rho_2,N_2 \ra^B_{\mathbb{T}_2^1}$ by performing an inverse Fourier transform of $| g,h \ra_{\mathbb{T}_2^1}$:
\begin{align} \nn
	| \rho_1 \rho_2,N_2 \ra^A_{\mathbb{T}^1_2} &=
	\frac{1}{|G|}\sum_{g,h}\sum_{\wro_1,\wro_2 \atop \{M,N\} \backslash N_2}
	\sqrt{d_{\rho_1}d_{\wro_1}}D^{\rho_1}_{M_1N_1}(g \smo \delta_h)
	\overline{D^{\wro_1}_{\widetilde{M}_1\widetilde{N}_1}(g^{-1}hg \smo \delta_{g^{-1}})} \\[-1em]
	\label{puncS1} 
	& \hspace{9em} \times
	\Big({}^{\, \rho_1^\ast \; \rho_1 \; \rho_2^\ast}_{M_1N_1N_2}\Big)
	\overline{\Big({}^{\, \wro_1^\ast \; \wro_1 \; \wro_2^\ast}_{\widetilde{M}_1\widetilde{N}_1\widetilde{N}_2}\Big)} \; |\wro_1\wro_2,\widetilde{N}_2 \ra^B_{\mathbb{T}^1_2} \; .
\end{align}
Let us now consider the quantity
\be
	\sum_{M_1,N_1}D^{\rho_1}_{M_1N_1}(g \smo \delta_h)
	\Big({}^{\, \rho_1^\ast \; \rho_1 \; \rho_2^\ast}_{M_1N_1N_2}\Big) \;.
\ee
The invariance of the Clebsch-Gordan coefficients allows to rewrite this expression as follows
\begin{align}
	\sum_{M_1,N_1 \atop O_1,P_1,P_2}\sum_{k_1,k_2}D^{\rho_1}_{M_1N_1}(g \smo \delta_h)
	D^{\rho_1^\ast}_{M_1O_1}(a \smo \delta_{k_1})D^{\rho_1}_{N_1P_1}(a \smo \delta_{k_2})
	{D^{\rho_2^\ast}_{N_2P_2}(a \smo \delta_{k_2^{-1}k_1^{-1}})}
	\Big({}^{\, \rho_1^\ast \, \rho_1 \, \rho_2^\ast}_{O_1P_1P_2}\Big) \;.
\end{align}
Using the defining property of representations and performing explicitly the multiplication, we obtain the following equality
\begin{align}
	\nn
	&\sum_{M_1,N_1}D^{\rho_1}_{M_1N_1}(g \smo \delta_h)
	\Big({}^{\, \rho_1^\ast \; \rho_1 \; \rho_2^\ast}_{M_1N_1N_2}\Big) \\ & \q = 
	\sum_{O_1,P_1,P_2}
	D^{\rho_1}_{O_1P_1}(a^{-1}ga \smo \delta_{a^{-1}ha})D^{\rho_2^\ast}_{N_2P_2}(a \smo \delta_{g^{-1}h^{-1}gh})
	\Big({}^{\, \rho_1^\ast \, \rho_1 \, \rho_2^\ast}_{O_1P_1P_2}\Big) 
	\label{puncS0}
\end{align}
Following exactly the same steps, we obtain another formula
\begin{align}
	\nn
	&\sum_{M_1,N_1}
	\overline{D^{\rho_1}_{M_1N_1}(g^{-1}hg \smo \delta_{g^{-1}})}
	\overline{\Big({}^{\, \rho_1^\ast \; \rho_1 \; \rho_2^\ast}_{M_1N_1N_2}\Big)} \\ & \q = 
	\sum_{O_1,P_1,P_2}
	\overline{D^{\rho_1}_{O_1P_1}(a^{-1}g^{-1}hga \smo \delta_{a^{-1}g^{-1}a})}
	\overline{D^{\rho_2^\ast}_{N_2P_2}(a \smo \delta_{g^{-1}h^{-1}gh})}
	\overline{\Big({}^{\, \rho_1^\ast \, \rho_1 \, \rho_2^\ast}_{O_1P_1P_2}\Big)}
	\label{puncS3}
\end{align}
Using \eqref{puncS0} and \eqref{puncS3}, the expression \eqref{puncS1} becomes
\begin{align}
	\nn
	|\rho_1\rho_2,N_2 \ra_{\mathbb{T}_2^1}^A 
	&=
	\frac{1}{|G|^2}\sum_{g,h,a}\sum_{\wro_1,\wro_2 \atop \{M,N\} \backslash N_2}
	\sqrt{d_{\rho_1}d_{\wro_1}}
	D^{\wro_2^\ast}_{N_2M_2}(a \smo \delta_{g^{-1}h^{-1}gh})
	\overline{D^{\rho_2^\ast}_{\widetilde{N}_2\widetilde{M}_2}(a \smo \delta_{g^{-1}h^{-1}gh})}
	\\[-0.5em] \nn & \hspace{8em} \times
	D^{\rho_1}_{M_1N_1}(a^{-1}ga \smo \delta_{a^{-1}ha})
	\overline{D^{\wro_1}_{\widetilde{M}_1\widetilde{N}_1}(a^{-1}g^{-1}hga \smo \delta_{a^{-1}g^{-1}a})} \\
	& \hspace{8em} \times
	\Big({}^{\, \rho_1^\ast \; \rho_1 \; \rho_2^\ast}_{M_1N_1M_2}\Big) 
	\overline{\Big({}^{\, \wro_1^\ast \; \wro_1 \; \wro_2^\ast}_{\widetilde{M}_1\widetilde{N}_1\widetilde{M}_2}\Big)}
	| \wro_1\wro_2,\widetilde{N}_2 \ra^B_{\mathbb{T}_2^1} \; .
\end{align}
Looking at \eqref{puncS0}, we see that that the delta functions encoding the flatness constraint associated to this fusion basis state is implicitly encoded in the contraction pattern of the representation matrices with the Clebsch-Gordan coefficients. Using such relations, our expression  becomes
\begin{align}
		\nn
	|\rho_1\rho_2,N_2 \ra_{\mathbb{T}_1}^A 
	&=
	\frac{1}{|G|^2}\sum_{g,h \atop a,b}\sum_{\wro_1,\wro_2 \atop \{M,N\} \backslash N_2}
	\sqrt{d_{\rho_1}d_{\wro_1}}
	D^{\wro_2^\ast}_{N_2M_2}(a \smo \delta_b)
	\overline{D^{\rho_2^\ast}_{\widetilde{N}_2\widetilde{M}_2}(a \smo \delta_b)}\\[-0.5em] 
	& \hspace{8em} \times \nn
	D^{\rho_1}_{M_1N_1}(g \smo \delta_{h})
	\overline{D^{\wro_1}_{\widetilde{M}_1\widetilde{N}_1}(g^{-1}hg \smo \delta_{g^{-1}})} \\
	& \hspace{8em} \times
	\Big({}^{\, \rho_1^\ast \; \rho_1 \; \rho_2^\ast}_{M_1N_1M_2}\Big) 
	\overline{\Big({}^{\, \wro_1^\ast \; \wro_1 \; \wro_2^\ast}_{\widetilde{M}_1\widetilde{N}_1\widetilde{M}_2}\Big)}
	| \wro_1\wro_2,\widetilde{N}_2 \ra^B_{\mathbb{T}_2^1} \; .
\end{align}
The orthogonality relation for the representations of the Drinfel'd double finally provides our definition
\begin{align}\nn
	| \rho_1\rho_2,N_2 \ra^A_{\mathbb{T}_2^1}
	& =\sum_{\widetilde{\rho}_1}S^{\rho_2}_{\rho_1\widetilde{\rho}_1}
	|\rho_1\rho_2,N_2 \ra_{\mathbb{T}_2^1}^B
\end{align}
with
\begin{equation*}
	S^{\rho_2}_{\rho_1\widetilde{\rho}_1}:=
	\frac{1}{|G|}\sum_{g,h}\sum_{\{M,N\}}
	\frac{\sqrt{d_{\rho_1}d_{\wro_1}}}{d_{\rho_2}}
	\overline{D^{\rho_1}_{N_1M_1}(g \smo \delta_h)}\;
	\overline{D^{\wro_1}_{\widetilde{M}_1\widetilde{N}_1}(h \smo \delta_g)}
	\Big({}^{\, \rho_1^\ast \; \rho_1 \; \rho_2^\ast}_{M_1N_1M_2}\Big) 
	\overline{\Big({}^{\, \wro_1^\ast \; \wro_1 \; \rho_2^\ast}_{\widetilde{M}_1\widetilde{N}_1M_2}\Big)} 
\; .\end{equation*}
In the case of a two-torus without punctures, the previous formula boils down to
	\begin{align} 
	\label{Smat}
	| \rho \ra_{\mathbb{T}_2}^{\rm A} 
	&\,=\, \sum_{\widetilde{\rho}}\bigg(\frac{1}{|G|}\sum_{g,h}
	\overline{\chi^{{\rho}}(g \smo \delta_h)}\;
	\overline{\chi^{\widetilde{\rho}}(h \smo \delta_g)}\bigg) \; | \widetilde{\rho}\ra_{\mathbb{T}_2}^{\rm B}
	=: \sum_{\widetilde{\rho}} S^{\rho\widetilde{\rho}}\;|\widetilde{\rho} \ra^{\rm B}_{\mathbb{T}_2} \; .
\end{align}

\section{Representations of the extended quantum triple algebra} \label{RepQT}

In this appendix, we define the representations of the algebra \eqref{5.3} that results from the gluing of two sliced open three-tori with curvature defects. Let us first introduce some notation: We define triples of conjugacy classes $T=(C_1, C_2, C_3, d)$ together with a degeneracy variable $d$. A triple is admissible if there are group elements $h_1 \in C_1, h_2 \in C_2$ and $h_3 \in C_3$ that satisfy
\begin{equation}
	\label{5.4p}
	h_3= h_2 h_1 h_2^{-1}h_1^{-1}  \; .
\end{equation}
The degeneracy index $d$ is only non-trivial, if---modulo the adjoint action---there is more than one such triple of group elements for a given triple $(C_1, C_2, C_3)$ of conjugacy classes. That is $d$ distinguishes between possibly multiple orbits of the adjoint action  in  $ {\cal C} \subset G\times G\times G$, where ${\cal C}$ is the subset satisfying \eqref{5.4}. We enumerate the  elements of the orbit associated to $T=(C_1, C_2, C_3, d)$ with $t_a \in {\cal C}$ and choose $t_1$ as a representative for this orbit. We denote by $Z_T$ the stabilizer group of $t_1$ under the adjoint action. The elements of the quotient $G/Z_T$ are denoted by $p_a$ such that $t_a=p_at_1 p^{-1}_a$.  We denote by $R$ an irreducible representation of the stabilizer group $Z_T$. An irreducible representation of the algebra \eqref{5.4p} is labeled by $(T,R)$ and its matrix elements are given by
\ba\label{5.5}
	D^{T,R}_{am,bn}(g\smo \delta_{h_1}\smo \delta_{h_2})  \,=\, \delta( h_1,  \text{First}(t_a)) \, \delta(h_2, \text{Sec}(t_a)) \, \delta (t_a, g \rhd t_b) \, D^R_{mn}( p_a^{-1} g p_b)  \; .
\ea
Here $\text{First}(t_a)$ extracts the first entry $h_1^a$ from the triple of group elements associated to $t_a$ and $\text{Sec}(t_a)$ the second, while $g \rhd t_b$ is the adjoint (diagonal) action on the triple $t_b$.  The verification that (\ref{5.5}) does define a representation of the algebra (\ref{5.3}) is straightforward and proceeds along the same line as for the Drinfel'd double representations or for the quantum triple representations in \cite{Delcamp:2017pcw}. Note finally that for the case where 
$C_3$ is the trivial conjugacy class, the representations defined here do reduce to the ones in \cite{Delcamp:2017pcw}.

\bibliographystyle{JHEP}
\bibliography{FB4D}

\end{document}